\documentclass[11pt]{article}
\pdfoutput=1 

\usepackage{graphicx}
\usepackage{calc}
\usepackage{rotating}
\usepackage[english]{babel}
\usepackage{graphicx}
\usepackage{subfig}
\usepackage{float}
\usepackage{amsmath}
\usepackage{amssymb}
\usepackage{amsthm}
\usepackage{latexsym}
\usepackage{dcolumn}
\usepackage{geometry}
\usepackage{color}
\usepackage{hyperref}
\usepackage{mciteplus}
\usepackage{slashed}
\usepackage{multirow}
\usepackage{hhline}

\def\gtwid{\mathrel{\raise.3ex\hbox{$>$\kern-.75em\lower1ex\hbox{$\sim
$}}}}
\def\vio{\mathrel{\hbox{$E$\kern-.60em\hbox{$/
$}}}}
\newcommand{\nn}{\nonumber}
\newcommand{\mueff}{\mu_{\mathrm{eff}}}

\newcommand{\hp}{{H^\pm}}

\newcommand{\ii}{\mathrm{i}}

\newcommand{\neut}[1]{\widetilde{\chi}^0_{#1}}

\newcommand{\hobs}{H_{\rm obs}}
\newcommand{\hsm}{H_{\rm SM}}

\newcommand{\lam}{\lambda}
\newcommand{\kap}{\kappa}

\newcommand{\akap}{A_\kappa}

\newcommand{\refeq}[1]{{eq.~(\ref{#1})}}

\newcommand{\be}{\begin{equation}}
\newcommand{\ee}{\end{equation}}

\begin{document}

\begin{center}
{\Large \bf {Dark Matter in the CP-violating NMSSM} \\
\vspace*{0.8cm}
{\large Waqas Ahmed$^a$, Mark Goodsell$^b$, Shoaib Munir$^c$ } \\[0.25cm]
{\small \sl $^a$ School of Mathematics and Physics, Hubei Polytechnic University, \\
Huangshi 435003, China} \\[0.25cm]
{\small \sl $^b$ Laboratoire de Physique Th\'eorique et Hautes Energies (LPTHE), \\
UMR 7589, Sorbonne Universit\'e et CNRS, 4 place Jussieu, 75252 Paris Cedex 05, France} \\ [0.25cm]
{\small \sl $^c$ East African Institute for Fundamental Research (ICTP-EAIFR), \\ 
University of Rwanda, Kigali, Rwanda} \bigskip \\
{\small \url{waqasmit@hbpu.edu.cn}, \url{goodsell@lpthe.jussieu.fr}, \url{smunir@eaifr.org}}}
\end{center}
\vspace*{0.4cm}


\begin{abstract}
\noindent
In the Next-to-Minimal Supersymmetric Standard Model there is a strong correlation between the mass terms corresponding to the singlet Higgs and the singlino interaction states, both of which are proportional to the parameter $\kappa$. If this parameter is complex, explicit CP-violation occurs in the Higgs as well as the neutralino sectors of the model at the tree level, unlike in the minimal scenario. A small magnitude of $\kappa$ typically yields a $\cal{O}$(10)\,GeV lightest neutralino with a dominant singlino component. In such a scenario, the phase of $\kappa$, beside modifying the properties of the five Higgs bosons, can also have a crucial impact on the phenomenology of the neutralino dark matter. In this study we perform a first investigation of this impact on the relic abundance of the dark matter solutions with sub-100\,GeV masses, obtained for parameter space configurations of the model that are consistent with a variety of current experimental data. 
\end{abstract}


\newpage
\section{Introduction}
\label{sec:intro}

Supersymmetric (SUSY) models with unbroken $R$-parity 
provide a viable candidate for the dark matter (DM) of the Universe, in the form of their lightest neutralino. The neutralinos are the mass eigenstates resulting from the mixing of the neutral fermionic superpartners of the electroweak (EW) gauge and Higgs bosons, and have Majorana masses. In the minimal superymmetric Standard Model (MSSM) \cite{Nilles:1983ge,Haber:1984rc}, which contains two complex Higgs super-multiplets, there are four neutralinos, $\neut{1,2,3,4}$. The interaction strengths of these neutralinos with Standard Model (SM) and SUSY particles are governed by their masses and compositions, i.e., the sizes of their \emph{bino}, \emph{wino} and \emph{higgsino} components.

The two scalar Higgs doublet fields of the MSSM yield a total of five Higgs states. In the limiting case when all the parameters in the Higgs and sfermion sectors are real, these states include two neutral scalars $h$ and $H$ (with $m_h < m_H$), a pseudoscalar $A$, and a charged pair $\hp$. In any model of new physics, (at least) one neutral scalar, which we generically refer to as the $\hsm$ here, ought to have properties consistent with those of the $\hobs$ discovered at the Large Hadron Collider (LHC)~\cite{Aad:2012tfa,CMS:2012qbp,Chatrchyan:2013lba}, i.e., a mass near 125\,GeV and SM-like coupling strengths. In the MSSM, maximising the tree-level mass of the lighter scalar $h$, which has an upper limit equal to the $Z$ boson mass, pushes the model into the so-called decoupling limit, where additionally its couplings to the vector bosons mimic those of the $\hobs$. Still, in order for $m_h$ to reach $\sim125$\,GeV, large loop corrections are needed from mainly the top quark, and its superpartners, the stops~\cite{Djouadi:2005gj,vanBeekveld:2019tqp}. As for the $\neut{1}$, its consistency with the Planck measurement of the DM relic abundance of the Universe, $\Omega_h^2$, for a mass below about 1\,TeV is only possible if it has a substantial bino component~\cite{Bergeron:2013lya,vanBeekveld:2016hug,Barman:2017swy,Roszkowski:2017nbc,KumarBarman:2020ylm,VanBeekveld:2021tgn}.

The Next-to-MSSM (NMSSM)~\cite{Fayet:1974pd,Ellis:1988er,Durand:1988rg,Drees:1988fc} (see, e.g.,~\cite{Ellwanger:2009dp,Maniatis:2009re} for reviews) is obtained by adding a Higgs singlet superfield in the MSSM, which results in raising the upper limit on the tree-level mass of the $\hsm$ in the model. This reduces the dependence of the $\hsm$ mass on the stop sector, and hence alleviates the fine-tuning problem to some extent. Due to the presence of the extra singlet superfield, the neutral Higgs sector of the NMSSM contains three scalars, $H_{1,2,3}$, and two pseudoscalars, $A_{1,2}$. Crucially, in the NMSSM there exists the possibility of the next-to-lightest CP-even Higgs boson, $H_2$, acting as the $\hsm$, with the lighter, sometimes even considerably so, $H_1$ still remaining undetected at the Large Electron Positron (LEP) collider as well as the LHC.   

The neutralino sector of the NMSSM also contains a fifth state which, when the lightest of all, can differ significantly from the $\neut{1}$ of the MSSM in its properties. In particular, in the above-mentioned scenario with the SM-like $H_2$, the $\neut{1}$ typically has a large singlino component, and a mass $\cal{O}$(10)\,GeV or even lower. Since the $H_1$, and generally also the $A_1$, lie below $\sim$100\,GeV~\cite{Bomark:2014gya,Ma:2020mjz}, this opens up multiple self-annihilation channels for the $\neut{1}$, which are precluded in the MSSM, in order to generate the correct $\Omega_h^2$~\cite{Abel:1992ts,Kozaczuk:2013spa,Cao:2013mqa,Han:2014nba,Ellwanger:2014dfa,Barman:2020vzm}. Potential LHC signatures of such a DM have been studied in~\cite{Cao:2011re,Das:2012rr,Ellwanger:2013rsa,Kim:2014noa,Ellwanger:2014hia,Han:2015zba,Potter:2015wsa,Enberg:2015qwa,Ellwanger:2016sur,Pozzo:2018anw,Domingo:2018ykx,Ellwanger:2018zxt,Abdallah:2019znp,Guchait:2020wqn,Wang:2020dtb}, and its detection prospects in~\cite{Ferrer:2006hy,Demidov:2010rq,Wang:2020xta} (see also \cite{Beck:2021xsv} for a phenomenological study of the DM in a two-Higgs doublet model with an additional singlet scalar).

Besides providing one of the leading candidates for low-mass DM, the NMSSM also entertains the possibility of explicit CP-violation in its Higgs sector at the tree level. This could serve as the additional source of CP-violation required for explaining the observed matter-antimatter asymmetry in the Universe through EW baryogenesis~\cite{Sakharov:1967dj,Cohen:1993nk,Quiros:1994dr,Rubakov:1996vz,Trodden:1998ym}. In the SM, the Cabibbo-Kobayashi-Maskawa (CKM) matrix is the lone and insufficient source of CP-violation, while the MSSM Higgs sector can only violate CP at higher orders~\cite{Pilaftsis:1998pe,Pilaftsis:1998dd,Pilaftsis:1999qt,Carena:2000yi,Choi:2000wz,Carena:2001fw,Carena:2002bb,Choi:2004kq,Frank:2006yh,Heinemeyer:2007aq,Carena:2015uoe,Mahmoudi:2018xml}. The CP-violating phases of the SUSY-breaking Higgs-sfermion-sfermion couplings, $A_{\tilde{f}}$, where $f$ denotes a SM fermion, in the MSSM can be radiatively transmitted to the Higgs sector, but are tightly constrained by the measurements of fermion electric dipole moments (EDMs)~\cite{Carena:2000ks,Abel:2001vy}. In the NMSSM, if the Higgs self-couplings, $\lambda$ and/or $\kappa$, appearing in the superpotential are complex, the scalar and pseudoscalar interaction eigenstates mix together to give five neutral CP-indefinite Higgs states; see \cite{Goodsell:2016udb,Grober:2017gut,Dao:2019qaz,Dao:2020dfb,Domingo:2021kud,Dao:2021khm} for recent studies of the NMSSM Higgs sector with CP violation and \cite{Slavich:2020zjv} for a review. We henceforth refer to this model as the cNMSSM. 

Several phenomenological scenarios emerging in the cNMSSM Higgs sector that are distinct from the NMSSM with real parameters (rNMSSM) have been studied in~\cite{Graf:2012hh,Moretti:2013lya,Munir:2013dya,King:2015oxa,Moretti:2015bua,Das:2017tob}. Importantly, the complex $\kappa$ parameter associated with the singlet superfield also appears in the entry of the neutralino mass matrix that corresponds to the singlino weak eigenstate. The impact of a non-zero phase of $\kappa$ on the phenomenology of the $\neut{1}$ DM has not been analysed in literature thus far. In this article, we take a first step in this direction, and investigate how the relic abundance of the $\neut{1}$ in the cNMSSM is affected by variations in this phase. We focus mainly on the (EW-scale) cNMSSM parameter space configurations that yield a sub-100\,GeV DM, which can be predominantly singlino-like. We also test the consistency of these solutions with the most important latest experimental constraints, including the Higgs boson data from the LHC and the electron and neutron EDMs, and study some of their phenomenological implications.

The article is organised as follows. In the next section we briefly revisit the Higgs and neutralino sectors of the cNMSSM. Section \ref{sec:numeric} contains details of our numerical analysis of the model's parameter space with the focus on the DM observables. In section \ref{sec:results} we present the results of our analysis, and we summarise our findings in section \ref{sec:concl}.


\section{\label{sec:sectors} The NMSSM with explicit CP-violation}

\subsection{\label{subsec:Higgs} The Higgs sector}

The superpotential of the NMSSM is written as
\begin{equation}
\label{Wpot}
W_{\rm NMSSM}\ =\ \widehat{U}^C {\bf h}_u \widehat{Q} \widehat{H}_u\:
+\:   \widehat{D}^C {\bf h}_d \widehat{H}_d \widehat{Q}  \: +\:
\widehat{E}^C {\bf h}_e \widehat{H}_d \widehat{L} \: + \:
\mu \widehat{H}_u \widehat{H}_d\  \: +\: 
\lambda \widehat{S} \widehat{H}_u \widehat{H}_d\  \: +\: 
\frac{\kappa}{3}\ \widehat{S}^3 \
\end{equation}
in terms of the singlet Higgs
superfield, $\widehat{S}$, besides the two $SU(2)_L$ doublet superfields,
 \begin{eqnarray}
\widehat{H}_u = \left(\begin{array}{c} \widehat{H}_u^+ \\ \widehat{H}_u^0
\end{array}\right)\,,\;
\widehat{H}_d = \left(\begin{array}{c} \widehat{H}_d^0 \\ \widehat{H}_d^-
\end{array}\right)\,,
\end{eqnarray}
of the MSSM. The above superpotential observes a discrete $Z_3$ symmetry, which is imposed in order to explicitly break the dangerous $U(1)_{PQ}$ symmetry, and renders it conformal-invariant by forbidding the $\mu \widehat{H}_u \widehat{H}_d$ term present in the MSSM superpotential. Here, the mixing between the $H_d^0$ field and the $H_u^0$ fields, necessary for each of them having a non-trivial vacuum expectation value (VeV) at the minimum of the potential, is instead generated by the $\lambda \widehat{S} \widehat{H}_u \widehat{H}_d$ term. This results in a dynamic $\mu_{\rm eff}\equiv \lambda s/\sqrt{2}$ term when the singlet field acquires a VEV, $s$, naturally near the SUSY-breaking scale.  

The tree-level Higgs potential of the NMSSM is obtained as
\begin{eqnarray}
\label{eq:Higgspot}
V_0 & = & \left|\lambda \left(H_u^+ H_d^- - H_u^0
H_d^0\right) + \kappa S^2 \right|^2 \nn \\
&&+\left(m_{H_u}^2 + \left|\lambda S\right|^2\right) 
\left(\left|H_u^0\right|^2 + \left|H_u^+\right|^2\right) 
+\left(m_{H_d}^2 + \left|\lambda S\right|^2\right) 
\left(\left|H_d^0\right|^2 + \left|H_d^-\right|^2\right) \nn \\
&&+\frac{g_1^2+g_2^2}{8}\left(\left|H_u^0\right|^2 +
\left|H_u^+\right|^2 - \left|H_d^0\right|^2 -
\left|H_d^-\right|^2\right)^2
+\frac{g_2^2}{2}\left|H_u^+ H_d^{0*} + H_u^0 H_d^{-*}\right|^2\nn \\
&&+m_{S}^2 |S|^2
+\big( T_\lambda \left(H_u^+ H_d^- - H_u^0 H_d^0\right) S + 
\frac{1}{3} T_\kappa\, S^3  + \mathrm{h.c.}\big)\,,
\end{eqnarray}
where $g_1$ and $g_2$ are the $U(1)_Y$ and $SU(2)_L$ gauge couplings. It is customary to define the trilinears proportional to the superpotential couplings as 
\begin{align}
  T_\lambda \equiv \lambda A_\lambda, \qquad T_\kappa \equiv \kappa A_\kappa.
\end{align}
$A_\lambda$ and $A_\kappa$ above are the soft SUSY-breaking counterparts of the superpotential couplings, and all of these can very well be complex parameters, with the corresponding phases, $e^{i\phi_{A_\lambda}}$, $e^{i\phi_{A_\kappa}}$, $e^{i\phi_\lambda}$, and $e^{i\phi_\kappa}.$ 

After spontaneous EW symmetry breaking, $V_0$ is evaluated at the vacuum, in terms of fields defined around their respective VEVs, $v_u$, $v_d$ and $s$, as %
\begin{eqnarray}
H_d^0&=&
\hphantom{e^{i\theta}}\,
\left(
\begin{array}{c}
\frac{1}{\sqrt{2}}\,(v_d+H_{dR}+iH_{dI}) \\
H_d^-
\end{array}
\right)\,, \nonumber \\[1mm]
H_u^0&=&
e^{i\theta}\,\left(
\begin{array}{c}
H_u^+\\
\frac{1}{\sqrt{2}}\,(v_u+H_{uR}+i H_{uI})
\end{array}
\right)\,, \\[1mm]
S&=&\frac{e^{i\varphi}}{\sqrt{2}}\,(s+S_R+iS_I)\,. \nonumber
\label{eq:higgsparam}
\end{eqnarray}

\noindent The potential then contains the phase combinations
\begin{center}
$\phi^\prime_\lambda-\phi^\prime_\kappa$ (with $\phi^\prime_\lambda \equiv \phi_\lambda+\theta+\varphi$ and $\phi^\prime_\kappa \equiv \phi_\kappa+3\varphi $), $\phi^\prime_\lambda+\phi_{A_\lambda}$, and $\phi^\prime_\kappa+\phi_{A_\kappa}$.
\end{center}
However, assuming vanishing spontaneous phases $\theta$ and $\varphi$, the last two phase combinations above can be determined up to a twofold ambiguity using the minimisation conditions of $V_0$, leaving $\phi^\prime_\lambda-\phi^\prime_\kappa$ as the only physical CP phase (see \cite{Munir:2013dya} for more details).

The potential $V_0$ with complex phases leads to a $5\times 5$ Higgs mass matrix, ${\cal M}_0^2$, in the ${\bf H}^T=(H_{dR},\,H_{uR},\,S_R,\,H_I,\,S_I)$ basis, with the
massless Nambu-Goldstone mode rotated away. After including the higher order corrections from various sectors of the model~\cite{Graf:2012hh,Munir:2013dya,Domingo:2015qaa}, the resulting Higgs mass matrix, ${\cal M}_H^2 = {\cal M}_0^2 + \Delta {\cal M}^2$, is diagonalised using an orthogonal matrix, $O$, as $O^T{\mathcal{M}}_0^2O={\rm diag}(m^2_{H_1}\;m^2_{H_2}\;m^2_{H_3}\;m^2_{H_4}\;m^2_{H_5})$. The masses of the five CP-mixed physical Higgs bosons thus obtained are ordered such that $m^2_{H_1}\leq m^2_{H_2}\leq m^2_{H_3} \leq m^2_{H_4} \leq m^2_{H_5}$.


\subsection{\label{subsec:neutralino} The Neutralino sector}

As noted in the Introduction, the fermion component of $\widehat S$, called the singlino,
mixes with the neutral gauginos, $\widetilde B^0$ and $\widetilde W_3^0$, and higgsinos, $\widetilde H_d^0$ and $\widetilde H_u^0$, to yield five neutralinos in the NMSSM.  The symmetric neutralino mass matrix in the gauge eigenstate basis, $\widetilde\psi^0 = (-\ii\widetilde B^0, -\ii\widetilde W_3^0, \widetilde H_d^0, \widetilde H_u^0, \widetilde S$), is written as
\begin{eqnarray}
\hspace*{0.2cm}
{\cal M}_{\widetilde{\chi}^0} =
\begin{pmatrix}
M_1 	& 0 	        &  -m_W \tan\theta_W \cos\beta      & m_W \tan\theta_W \sin\beta & 0  	\\
 0	& M_2 	& m_W \cos\beta                              &-m_W   \sin\beta               & 0  	\\
 -m_W \tan\theta_W \cos\beta  	&   m_W \cos\beta  	& 0				& -\mueff				& -\lambda v_u \\
 m_W \tan\theta_W \sin\beta	&  -m_W   \sin\beta   	& 	-\mueff			& 0				& -\lambda v_d \\
 0  	&  0	& 		-\lambda v_u		& 	 -\lambda v_d 			& 2\kappa s
\end{pmatrix}\,,
\label{eq:massmatrix}
\end{eqnarray}
with $m_W$ and $\theta_W$ being the $W$-boson mass and the weak mixing angle, respectively. 
The neutralino masses and compositions at the tree level thus depend
on the Higgs-sector parameters $\lam$, $\kap$, $\mueff$, $v_u$, $v_d$ and the
gaugino masses $M_1$ and $M_2$. When any of these parameters is complex, the mass matrix in \refeq{eq:massmatrix} can be diagonalised by a unitary matrix $N$, to give
$D=\text{diag}(m_{\neut{i}}) = N^* {\cal M}_{\widetilde{\chi}^0} N^\dag$,
for $i=1 - 5$. The neutralino mass eigenstates are then given by $\neut{i} = N_{ij} \widetilde\psi^0_j$, and are again ordered as $m_{\neut{1}}\leq m_{\neut{2}} \leq m_{\neut{3}} \leq m_{\neut{4}} \leq m_{\neut{5}}$. 

The $\neut{1}$, which is the lightest neutral SUSY particle and hence a DM candidate, is given by the linear combination
 \begin{eqnarray}
 \neut{1} = N_{11} \widetilde B^0 + N_{12} \widetilde W_3^0 + N_{13} \widetilde
 H_d^0 + N_{14} \widetilde H_u^0 + N_{15} \widetilde S^0.
 \end{eqnarray}
Thus, the relative sizes of the soft gaugino masses $M_{1,2}$, the $\mu_{\rm eff}$-parameter and the $\kappa s$ term determine whether the $\neut{1}$ is gaugino-, higgsino- or singlino-like. For example, in the limit $\mueff \ll \min[M_1,M_2]$, the term $[\mathcal{M}_{\widetilde{\chi}^0}]_{55} = 2\kap s = 2\frac{\kap \mueff}{\lam}$ in eq.~(\ref{eq:massmatrix}) results in a singlino-dominated $\neut{1}$ for $2\kap / \lam < 1$. Importantly, in SUSY models the charged higgsinos ($\widetilde H_u^+$ and $\widetilde H_d^-$) and winos ($\widetilde{W}^+$ and $\widetilde{W}^-$) also mix to form the chargino eigenstates, $\widetilde{\chi}^\pm_a~ (a=1,2)$. The mass matrix for the charginos is given by
\begin{eqnarray}
{\cal M}_{\widetilde{\chi}^\pm} =
\begin{pmatrix}
M_2 	& \sqrt{2}m_W \sin\beta	 	\\
\sqrt{2}m_W \cos\beta	& \mu_{\rm eff} 	\\
 \end{pmatrix}\,.
\end{eqnarray}
This implies that $M_2$ and $\mu_{\rm eff}$ have a lower bound of about 100\,GeV, owing to the non-observation of a chargino at the LEP collider. Since no such constraint exists on the $M_1$ and $2\kappa s$ terms, the $\neut{1}$ in the NMSSM can have a mass much lower than 100\,GeV as long as it is predominantly bino- and/or singlino-like. The presence of a certain amount of higgsino is, however, necessary to obtain a realistic relic abundance. 

The existence of the singlino in the NMSSM, even in the absence of the CP-violating phases noted above, leads to some unique possibilities in the context of DM phenomenology, compared to the MSSM. In the limit of large $\tan\beta\equiv v_u/v_d$ and large $m_{A}\equiv\frac{\lambda s}{\sin 2\beta}(\sqrt{2}A_\lambda+ ks$) (which effectively decouples the doublet-like $H_3$ and $A_2$ from the rest of the particle spectrum, so that $m_{A_2}\simeq m_A$), the masses of the two lightest CP-even scalars can be approximated by~\cite{Miller:2003ay}
\begin{eqnarray}
\label{eq:mh12form}
M_{H_1,H_2}^2&\approx&\frac{1}{2}\left\{m_Z^2+4(\kappa s)^2+\kappa s
  A_\kappa \right. \nonumber \\ 
  && \left.\sqrt{\left[m_Z^2-4(\kappa s)^2-\kappa s
  A_\kappa\right]^2+4\lambda^2 v^2\left[2\lambda s-\left(A_\lambda+\kappa s\right)\sin 2\beta\right]^2}\right\}\,, 
\end{eqnarray}
where $v\equiv\sqrt{v_u^2+v_d^2}$. Thus the mass of the lighter of these two (when the heavier one is required to be the $\hsm$) scales with $\kappa s$, as does that of the singlino. At the same time, the mass-squared of the lighter pseudoscalar, which is almost purely a singlet, reduces to 
\begin{equation}
\label{eq:ma1}
m_{A_1}^2 \simeq -3\kappa s A_\kappa.
\end{equation}
This correlation between the masses of the $A_1$ and the $\neut{1}$ implies that they can be naturally close to each other, thus opening the possibility of the former's self-annihilation via the latter. Evidently, while $H_1$ can also have a mass in the vicinity of $m_{A_1}$, it is more strongly constrained by Eq.~(\ref{eq:mh12form}) from taking values close to $2\neut{1}$. Even if it does acquire the correct mass, the $\neut{1}$ annihilation via $s$-channel $H_1$ is $p$-wave suppressed, which would make its consistency with the thermal relic abundance difficult. 

When the CP-violating phases of $\kappa$ and $\lambda$ are turned on, they enter the tree-level neutralino mass matrix independently of each other, unlike the combination $\phi^\prime_\lambda-\phi^\prime_\kappa$ of the Higgs sector. In addition, $M_1$ and $M_2$ can also be complex parameters, which would be radiatively induced into the Higgs sector at higher orders. Given the composition of the $\neut{1}$, the size(s) of the most relevant phase(s) would then affect not only its physical mass, but also its interaction strengths with other particles. Here, our focus on a sub-100\,GeV DM, which is also preferably singlino-like (since low-mass bino-like solutions exist in the MSSM too and have been extensively studied), makes $\phi_\kappa$ the most obvious choice to investigate the impact of.

\subsection{\label{subsec:EDMs} The electric dipole moments of fermions}

Beyond the Born approximation, various CP-violating phases are (co-)induced in the Higgs and neutralino sectors of the cNMSSM. Such phases are subject to constraints from the non-observation of the EDMs of the electron and the neutron. The most recent limits on these EDMs read

\begin{equation}\label{eq:edm-limits}
|d_e| < 1.1\times 10^{-29}\,e\,{\rm cm}\,\cite{ACME:2018yjb};~~~~|d_n| < 1.8\times 10^{-26}\,e\,{\rm cm}\,\cite{nEDM:2020crw}.
\end{equation}

The limit on the electron EDM above is based on the thorium monoxide experiment, and is more stringent than the one from the HfF$^+$ experiment~\cite{Cairncross:2017fip}. In SUSY models, the one-loop EDMs of the charged leptons and the light quarks are induced by chargino and neutralino exchange diagrams. These should in principle constrain the CP-violating phases of $M_1$, $M_2$, $\lambda$ and $\kappa$, appearing in the chargino/neutralino sectors at the tree level. In the SUSY spectrum generator code used for our analysis, details of which will be provided in the next section, calculation of the one-loop contributions to $d_e$ and $d_n$, as well as to $d_\tau$, in the cNMSSM is currently implemented. Note that additional constraints also come from mercury~\cite{Griffith:2009zz} and thallium~\cite{Regan:2002ta} EDMs, but these can generally be evaded if the masses of the first two generations of squarks are taken to be sufficiently heavy, as discussed in~\cite{Goodsell:2016udb}.

At the two-loop level, the Higgs-mediated Barr-Zee type diagrams can also contribute significantly to the electron and neutron EDMs. However, several studies have shown that even when these two-loop effects are taken into account, the phase $\phi^\prime_\kappa$ is very weakly constrained by the fermionic EDMs~\cite{Graf:2012hh,Cheung:2010ba,Cheung:2011wn,King:2015oxa}, especially for smaller values of $|\kappa|$. This is in contrast with the other phases, especially $\phi_{A_{\tilde{f}}}$ (the phases of the Higgs-sfermion-sfermion trilinear couplings), which enter the Higgs sector at the one-loop level. Therefore, besides the reason noted above, we choose the $\phi^\prime_\kappa$ as the sole representative CP-violating phase additionally to minimise the potential impact of these two-loop diagrams, which are not accounted for in our numerical code. Implementation of the complete set of contributions to the EDMs in our numerical code would go beyond the scope of this article, which aims to explore the DM properties when a non-zero phase appears in the neutralino sector.

\section{\label{sec:numeric} Numerical Analysis}

The radiative corrections to the tree-level Higgs and neutralino mass matrices make the parameters of the other model sectors highly relevant also. However, on the one hand, the NMSSM with grand-unification-inspired boundary conditions is very tightly constrained by the current experimental results, and on the other hand, the most general NMSSM contains more than a hundred free parameters defined at the EW scale. Thus, in order to draw inferences for a particular sector of the model, it is imperative to make multiple assumptions about the free parameters that only impinge at higher orders.

For our numerical analysis, we therefore adopted the following (universality) conditions to impose on the parameter space of the $Z_3$-symmetric cNMSSM at the EW scale:
\begin{gather}
M_{\tilde{f}} \equiv M_{Q_{1,2,3}} = M_{U_{1,2,3}} = M_{D_{1,2,3}} =
M_{L_{1,2,3}} = M_{E_{1,2,3}}\,, \nonumber \\
T_{\tilde{f}} \equiv T_{\tilde{t}} = T_{\tilde{b}} = T_{\tilde{\tau}}\,, \nonumber
\end{gather}
where $M^2_{Q_{1,2,3}},\,M^2_{U_{1,2,3}},\,M^2_{D_{1,2,3}},\,M^2_{L_{1,2,3}}$ and $M^2_{E_{1,2,3}}$ are the squared soft masses of the sfermions. The (less-often used) parameter $T_{\tilde{f}}$ corresponds to sfermion trilinear couplings; usually these are taken to be proportional to the Yukawas, such that $T_{\tilde{t}}^{ij} = Y_u^{ij} A_{\tilde{t}},$ where $i,j$ are generation indices. In our numerical code, we specified $T_{\tilde{f}}$ directly at the low scale. We fixed all the elements of $T_{\tilde{f}}$ to small values ($1$\,GeV for the diagonal terms and zero otherwise), except for $T_{\tilde{f}}^{(3,3)} = T_{\tilde{t}}^{(3,3)} = T_{\tilde{b}}^{(3,3)} = T_{\tilde{\tau}}^{(3,3)}$, which we left as a free parameter to be scanned over an extended range. The reason for this was to increase the probability of the consistency of $m_{H_{\rm SM}}$, the dominant corrections to which increase proportionally to $(A_{\tilde{t}} - \mueff\cot\beta)^2$, with $m_{H_{\rm obs}}$. We likewise scanned over wide ranges of $M_1$ and $M_2$ to allow maximal possible variations in the $\neut{1}$ composition. On the other hand, $M_{\tilde{f}}$ and $M_3$ were fixed to sufficiently large values of 2\,TeV and 3\,TeV, respectively, so that the sfermions and the gluino could evade the direct search limits from the LHC.

As for the CP-violating phases, in light of the discussion in the previous section, we fixed $\phi^\prime_\lambda = \phi_{M_1} = \phi_{M_2} = \phi_{T_{\tilde{f}}} = \varphi = \theta = 0$ (so that $\phi^\prime_\kappa = \phi_\kappa$). However, the quantities that we choose for the solution of the (five independent) tadpole equations are
$$
m_{H_d}^2,\,m_{H_u}^2,\,m_S^2,\,\mathrm{Im}(T_\kappa),\,\mathrm{Im}(T_\lambda)\,.
$$
The first three of these are standard choices familiar from the rNMSSM. However, once we break CP we have two additional non-trivial tadpole equations that must be satisfied, and it is logical to choose the complex part of the trilinear parameters, since these lead to the smallest impact on the spectrum and their magnitude will only be proportional to the violation of CP. With the only non-zero CP-violating phase being $\phi_\kappa$, this leads to
\begin{align}
  \mathrm{Im} (T_\lambda) =& \mu_{\rm eff} \mathrm{Im} (\kappa) - \frac{\sqrt{2}}{s v_u} \mathrm{Re}\left( \frac{\partial \Delta V_0}{\partial H_{dI}} \right), \nn\\
  \mathrm{Im} (T_\kappa) =& \frac{3 \lambda v_d v_u }{\sqrt{2} s} \mathrm{Im} (\kappa) + \frac{\sqrt{2}}{s^3} \mathrm{Re}\left( s \frac{\partial \Delta V_0}{\partial S_{I}} - v_d  \frac{\partial \Delta V_0}{\partial H_{dI}}\right).
\end{align}
As briefly noted in Sec.~\ref{subsec:Higgs}, the real parts of these trilinears are fixed as inputs:
\begin{align}
  \mathrm{Re}(T_\lambda) =& \mathrm{Re}(\lambda A_\lambda), \nn\\
  \mathrm{Re}(T_\kappa) =& \mathrm{Re}(\kappa A_\kappa),
\end{align}
where now $A_\lambda, A_\kappa$ are taken to be real. This means that at the tree level both the trilinear couplings pick up phases from the phase of $\kappa$ (which are, however, small for  $A_\lambda \gg \mu_{\rm eff}$ and $A_\kappa \gg v_d v_u/s$, and are modified at the higher orders). 

To generate the particle spectrum for a given configuration of the final set of the free parameters,
\begin{center}
$|M_1|$\,, $|M_2|$\,, $|T_{\tilde{f}}|$, $\tan\beta$\,, $|\lambda|$\,, $|\kappa|$\,, $\mu_{\rm eff}$\,, $A_\lambda$\,, $A_\kappa$\,, $\phi_\kappa$\,,
\end{center}
we incorporated the cNMSSM into the public fortran code {\tt SPheno-v4.0.4}~\cite{Porod:2003um,Porod:2011nf} using the Mathematica package {\tt SARAH-v4.14.4}~\cite{Staub:2008uz,Staub:2013tta,Goodsell:2014bna,Goodsell:2015ira,Staub:2015kfa,Goodsell:2016udb,Braathen:2017izn}.\footnote{We note that {\tt NMSSMTools} provides a spectrum generator capable of handling CP violation \cite{Domingo:2015qaa} that contains a less accurate computation of the Higgs masses; whereas {\tt NMSSMCALC} \cite{Baglio:2013iia} provides an equivalent computation of the Higgs masses \cite{Dao:2021khm}.} Besides the mass spectrum, {\tt SPheno} also computes the decay widths and branching ratios (BRs) of the Higgs and SUSY particles (at one loop for CP-conserving models \cite{Goodsell:2017pdq}, but at leading order only for CP-violating ones), as well as a multitude of flavour and other low-energy observables. We linked {\tt SPheno} with the public program {\tt MultiNest-v3.7}~\cite{Feroz:2008xx} for generating output files for sampled configurations of the free parameters from their defined ranges. Multiple scans were performed, with each one corresponding to $\phi_\kappa$ fixed to one of the five selected values: $0^\circ$ (the CP-conserving case, i.e., the rNMSSM but with the five Higgs bosons ordered by their masses, irrespective of their CP-identities), $30^\circ,\,60^\circ,\,135^\circ$, and $180^\circ$.\footnote{We ignored $\phi_\kappa > \pi$, since we expected the real part of $\kappa$ to be dominant by far, and hence the overall behaviors of the calculated observables to be approximately symmetric around $\phi_\kappa =\pi$. This was nevertheless verified numerically for the sample value of $\phi_\kappa = 300^\circ$.} In order to calculate $\Omega_{\neut{1}}{h^2}$ and other DM observables for each sampled parameter space point, we also produced a {\tt CalcHEP}~\cite{Belyaev:2012qa} model file for the cNMSSM with {\tt SARAH}, which was then embedded in the public code {\tt MicrOmegas-v5.2.4}~\cite{Belanger:2006is,Belanger:2014vza,Barducci:2016pcb}. 

\begin{table}[tbp]
\begin{center}
\begin{tabular}{||c||c|c|c|c|c||} \hline \hline
  Parameter  & $M_1$\,(GeV)  &  $M_2$\,(GeV)  & $T_{\tilde{f}}$\, (GeV)  & $\tan\beta$	& 	 \\ \hline
  Range &    [1, 1000]	&[100, 2000]	& [$-7000$, $-2000$]	& [1, 20]  &\\ \hline\hline
  Parameter  & $\lambda$ 	&  $\kappa$ 		&   $\mu_{\rm eff}$\,(GeV) 	&$A_\lambda$\,(GeV)  	&$A_\kappa$\,(GeV) \\ \hline
  Range &[0.1, 0.7]  & [0.001, 0.3] &[100, 500] &[$500$, $3000$] &[$\mp 500$, 0] \\ \hline\hline
\end{tabular}
\caption{Scanned ranges of the cNMSSM parameters. Seperate scans were run for  $\phi_\kappa$ chosen from $\{0^\circ,30^\circ,60^\circ,135^\circ,180^\circ\}$.}
\label{tab:params}
\end{center}
\end{table}

The purpose of these scans was to find parameter space points for which
\begin{enumerate}
\item Either $H_2$ or $H_3$ had a mass in the $122-128$\,GeV range (thus allowing a theoretical uncertainty of $\pm 3$\,GeV around the - assumed - experimental central value of $m_{H_{\rm SM}}=125$\,GeV). This implied that there would at least be one light Higgs boson available for potential $s$-channel annihilation of the DM.

\item The $\hsm$ (whether $H_2$ or $H_3$) had the $\gamma\gamma$, $ZZ$, $\tau\tau$ and $b\overline{b}$ effective couplings lying within $\pm 0.2$ units of the SM expectation of 1.\footnote{While it is in principle possible to constrain these couplings using the latest combined measurements from the LHC (e.g.,~\cite{Langford:2021osp}) using a program like {\tt HiggsSignals-2}~\cite{Bechtle:2020uwn} instead we chose (for simplicity) to allow up to 20\% deviation from the SM values. The figure of 20\% corresponds roughly to the experimentally quoted uncertainties (we also checked that using a smaller value negligibly impacted our results); we were concerned that a combination was overly pessimistic regarding finding valid points, and not primarily concerned with tweaking the (heavier) $\hsm$ which otherwise plays little role in the dark matter properties.}

\item The theoretical predictions of the following $B$-physics observables lied within 2$\sigma$ deviation from their quoted experimental values. 
\begin{itemize}
\item ${\rm BR}(B\to X_s \gamma) \times 10^{4} = 3.32\pm0.15$~\cite{HFLAV:2016hnz},
\item ${\rm BR}(B_u\to \tau^\pm \nu_\tau) \times 10^{4} = 1.06\pm0.19$~\cite{HFLAV:2016hnz}, 
\item ${\rm BR}(B_s \to \mu^+ \mu^-)\times 10^{9} = 3.0\pm 0.85$~\cite{Aaij:2017vad}. 
\end{itemize}

\item None of the five Higgs states were excluded by the limits from the LEP, TeVatron and LHC searches implemented within the program {\tt HiggsBounds-v5.7.0}~\cite{Bechtle:2020pkv}.

\item The $|d_e|$ and $|d_n|$ satisfied the experimental upper bounds given in \refeq{eq:edm-limits}.

\item The relic abundance of the $\neut{1}$ never exceeds $+10\%$ of the Planck measurement of $\Omega h^2= 0.119$~\cite{Planck:2018vyg}. This allowance in $\Omega_{\neut{1}}{h^2}$ is to crudely account for the rather large uncertainty in its theoretical estimation, due to the higher order corrections in SUSY models \cite{Baro:2007em,Boudjema:2011ig,Beneke:2014gla,Belanger:2016tqb,Harz:2016dql,Schmiemann:2019czm,Branahl:2019yot}. Note that {\tt MultiNest} performs a multimodal sampling of a model's parameter space based on Bayesian evidence estimation. Our purpose for using this package was simply to scan the parameter space in a more efficient way than random sampling, rather than to draw Bayesian inferences about it. To this end, we defined a Gaussian likelihood function with a peak at $\Omega_{\neut{1}}{h^2}= 0.119$ and a width of $\pm 10\%$ of this value in {\tt MultiNest}. Evidently, the scan collected a number of points far away from the peak also. From these, we removed all the points with $\Omega_{\neut{1}}{h^2} > 0.131$, but retained also the ones for which $\Omega_{\neut{1}}{h^2} < 0.107$ so as to accommodate alternative possibilities, such as non-thermal $\neut{1}$ production~\cite{Acharya:2008bk} or multi-component DM~\cite{Zurek:2008qg}.

\end{enumerate}

\begin{table}[tbp]\centering
  \raisebox{.6\height}{
   \begin{tabular}{|c|c|c|c|c|} \hline\hline
Parameter & TP1 & TP2 & TP3 & TP4 \\ \hline\hline 
$M_1$ (GeV) & 42 & 386 & 689 & 602 \\ 
$M_2$ (GeV) & 1967 & 1565 & 1273 & 773 \\ 
$T_{\tilde{f}}$ (GeV) & -2688 & -3414 & -3938 & -4132 \\ 
$\tan \beta$ & 13.69 & 10.80 & 14.82 & 6.84 \\ 
$\lambda$ & 0.227 & 0.323 & 0.659 & 0.252 \\ 
$|\kappa|$ & 0.178 & 0.012 & 0.003 & 0.016 \\ 
$\mu_{\rm eff}$ & 458 & 252 & 192 & 322 \\ 
$A_\lambda$ (GeV) & 2694 & 2613 & 2904 & 2165 \\ 
$A_\kappa$ & -11.70 & -2.44 & -57.45 & -31.06 \\ 
\hline\hline
          \end{tabular}} \begin{tabular}{|c|c|c|c|c|} \hline\hline
Parameter & TP1 & TP2 & TP3 & TP4 \\ \hline\hline 
$\phi_\kappa$ (degrees) & 30 & 30 & 30 & 30 \\ 
$\Omega h^2$ & 0.130 & 0.127 & 0.119 & 0.079 \\ 
$m_{\tilde{\chi}_1^0}$ (GeV) & 41 & 20 & 6 & 43 \\ 
$m_{\tilde{\chi}_2^0}$ (GeV) & 463 & 251 & 216 & 317 \\ 
$m_{\tilde{\chi}_3^0}$ (GeV) & 469 & 265 & 228 & 333 \\ 
$m_{\tilde{\chi}_4^0}$ (GeV) & 723 & 393 & 688 & 603 \\ 
$m_{\tilde{\chi}_5^0}$ (GeV) & 1973 & 1578 & 1292 & 804 \\ 
$m_{\tilde{\chi}_1^\pm}$ (GeV) & 466 & 256 & 193 & 323 \\ 
$m_{\tilde{\chi}_2^\pm}$ (GeV) & 1988 & 1597 & 1310 & 813 \\ 
$m_{h_1}$ (GeV) & 87 & 16 & 14 & 45 \\ 
$m_{h_2}$ (GeV) & 125 & 45 & 69 & 52 \\ 
$m_{h_3}$ (GeV) & 720 & 124 & 127 & 123 \\ 
$m_{h_4}$ (GeV) & 4302 & 2671 & 2970 & 2196 \\ 
$m_{h_5}$ (GeV) & 4302 & 2672 & 2974 & 2196 \\ 
$m_{H^\pm}$ (GeV) & 4303 & 2651 & 2830 & 2188 \\ 
\hline\hline
          \end{tabular} 
 \caption{\label{TAB:benchmarks} The input parameters and the spectra for the four selected test points.}
\end{table}


\section{\label{sec:results} Low-mass DM in the cNMSSM}

The scanned ranges of the nine free parameters (after fixing $\phi_\kappa$) of the cNMSSM are given in Table~\ref{tab:params}. Evidently, these ranges cannot entail all possible configurations. They are, however, guided by some previous studies~\cite{Bomark:2014gya,Han:2015zba,Enberg:2015qwa}, and are extensive enough to fulfill the necessary conditions for this analysis, i.e., of yielding a $H_2$ or $H_3$ with a mass around 125\,GeV, and a sub-100\,GeV singlino/bino-dominated $\neut{1}$. The upper cutoffs on the values of $\lam$ and $\kap$ are imposed to avoid the Landau pole. $A_0$ can in principle be both positive and negative, with a marginally different impact on the physical mass of the SM-like Higgs boson for an identical set of other input parameters in each case. Our purpose for using only its negative range was to enhance the efficiency of the numerical scanning code. Note that, at the EW scale $\kappa$ and $\akap$ are conventionally taken to be $>0$ and $<0$, respectively, in order to prevent negative mass-squared of the lightest pseudoscalar -- see Eq.\.({\ref{eq:ma1}). But here, for $90^\circ < \phi_\kappa < 180^\circ$, the real part of $\kappa$ becomes negative, and hence $\akap$ ought to be positive. Thus in the scans corresponding to $\phi_\kappa =135^\circ$ and $180^\circ$, $\akap$ was scanned over positive values only. 

\begin{figure}[tbp]
\begin{tabular}{cc}
\vspace*{-1.0cm}\hspace*{-0.9cm}\includegraphics*[width=9.5cm]{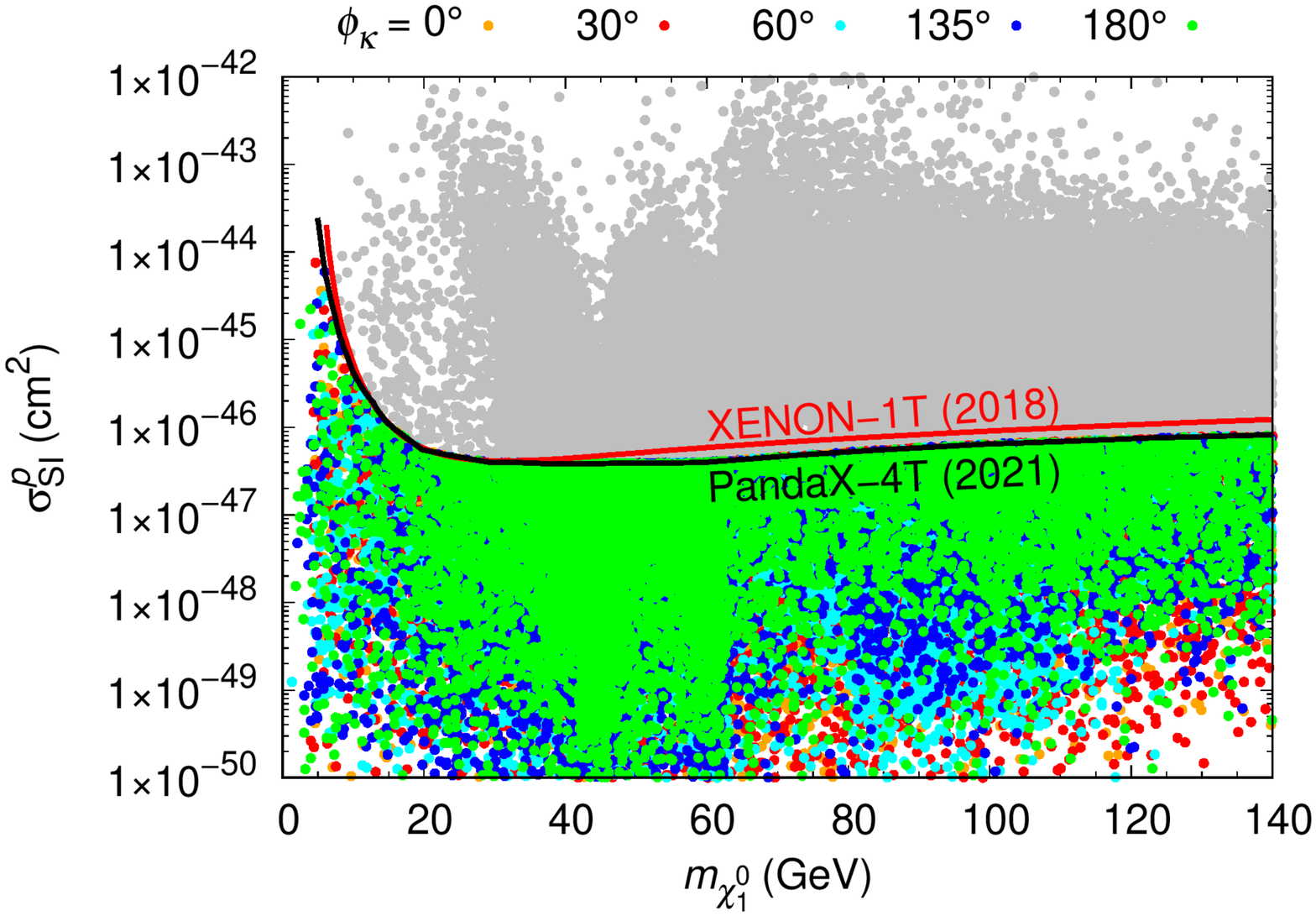} &
\hspace*{-1.2cm}\includegraphics*[width=9.5cm]{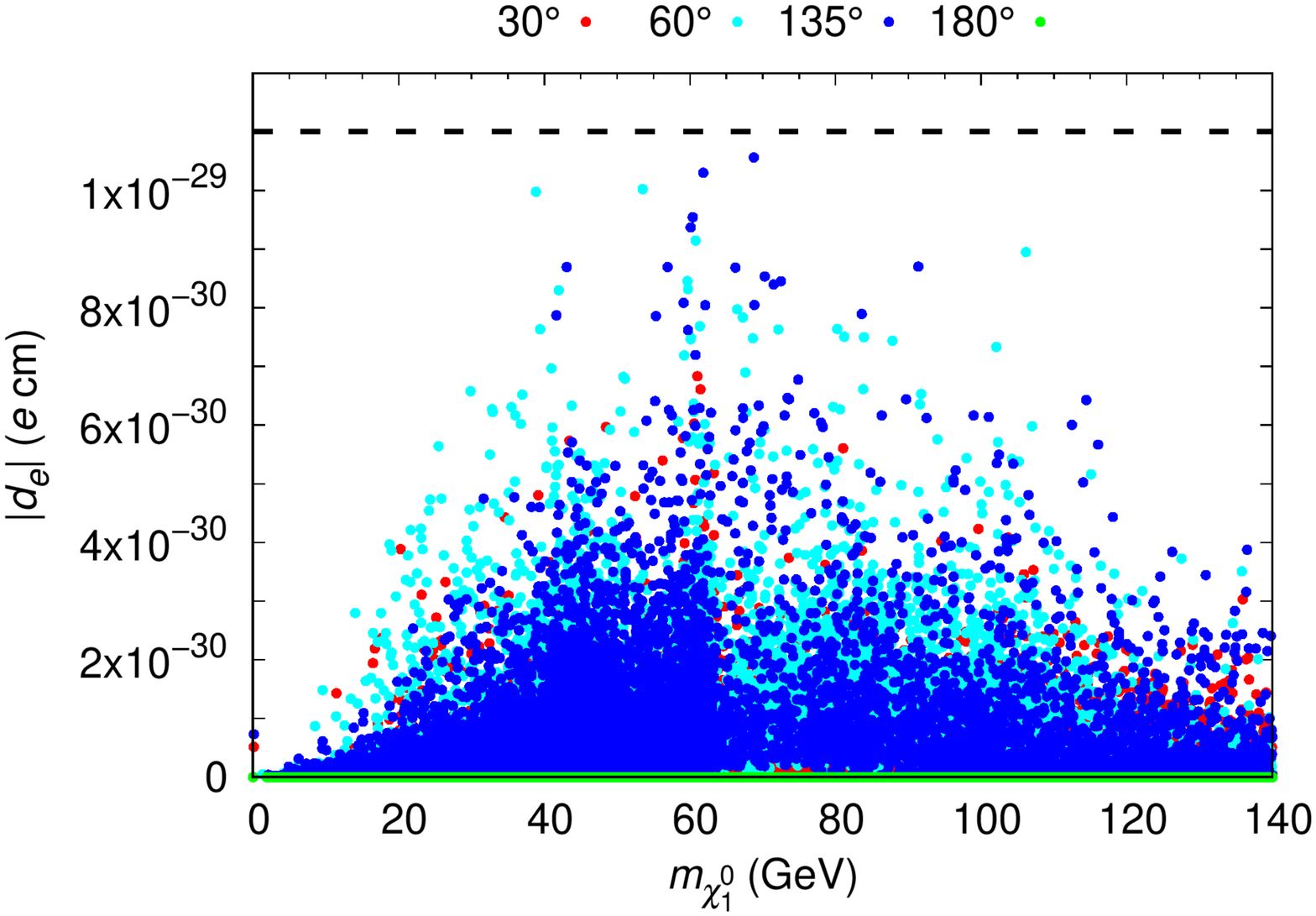}\\
\end{tabular}
\caption{\label{fig:DD-EDM} The spin-independent DM-proton cross section (left) and the electron EDM (right) as functions of the DM mass. Different colours of the points illustrate different values of $\phi_\kappa$, and the solid and dashed lines in the left and right panels, respectively, correspond to the experimental limits.}
\end{figure}

The left panel of Fig.~\ref{fig:DD-EDM} shows that a large number of parameter space points meeting all the conditions outlined in the previous section are ruled out by the latest limits on the cross section of the spin-independent DM-proton scattering, $\sigma^p_{SI}$, from the PandaX-4T Cmmissioning Run~\cite{PandaX-4T:2021bab} (for comparison, the most recent exclusion contour from the XENON-1T experiment~\cite{XENON:2018voc} is also shown). The right panel confirms the fact that $\phi_\kappa$ is indeed very weakly constrained by the one-loop contributions to $|d_e|$, since for almost all the successful scanned points its model prediction lies well below the experimental bound. 

In Fig.~\ref{fig:0-180} the $\neut{1}$ relic abundance is plotted as a function of its mass, separately for points from each scan with fixed $\phi_\kappa$. In this figure, the grey points in the background are the ones excluded by the PandaX-4T limits, and the coloured points further satisfy the following two conditions.
\begin{itemize}
\item The total invisible BR of the $\hsm$ is required to lie below the latest upper limit of 14.5\% from ATLAS \cite{ATLAS:2022yvh} (the corresponding limit from CMS \cite{CMS:2022qva} is slightly weaker). This BR accounted for, besides $\hsm \to \neut{1}\neut{1}$, decays like $\hsm \to \neut{i}\neut{1}$ ($i=2-5$), which are followed by $\neut{i} \to H_1\neut{1}/H_2 \neut{1}$ and subsequently $H_1/H_2 \to \neut{1}\neut{1}$. In addition, BRs for $\hsm \to H_1H_1/H_1H_2/H_2H_2$, followed by $H_1/H_2 \to \neut{1}\neut{1}$, which would also yield a 4$\neut{1}$ final state, were also included. For all the good points from our scans, however, the total invisible BR of $\hsm $ is by far dominated by the $\neut{1}\neut{1}$ decay.

\item Various searches targeting $\widetilde{\chi}^\pm_1$ and $\neut{i}$ production, such as \cite{ATLAS:2019lff,ATLAS:2020pgy,ATLAS:2019wgx}, are relevant here, since these states can decay to our (very) light $\neut{1}$ along with $W/Z/\hobs$. These analyses quote limits of up to $750$ GeV for winos with specific channels of decay. Searches for higgsinos are notoriously difficult due to their small production cross-section, and the limits from them are thus much weaker. The very latest ATLAS search \cite{ATLAS:2021moa} quotes limits of up to 210 GeV on a higgsino. A full recasting of these results using  {\tt MadAnalysis} \cite{Conte:2012fm,Conte:2014zja,Dumont:2014tja,Conte:2018vmg} for all our good points will go beyond the scope of this study. We nevertheless imposed the simplistic requirement that the higgsino (wino) component of a given $\neut{i}$ is less than 90\% if it is lighter than 210\,GeV (750\,GeV).
\end{itemize}

\begin{figure}[t!]
\begin{tabular}{cc}
\vspace*{-1.5cm}\hspace*{-0.9cm}\includegraphics*[width=9.5cm]{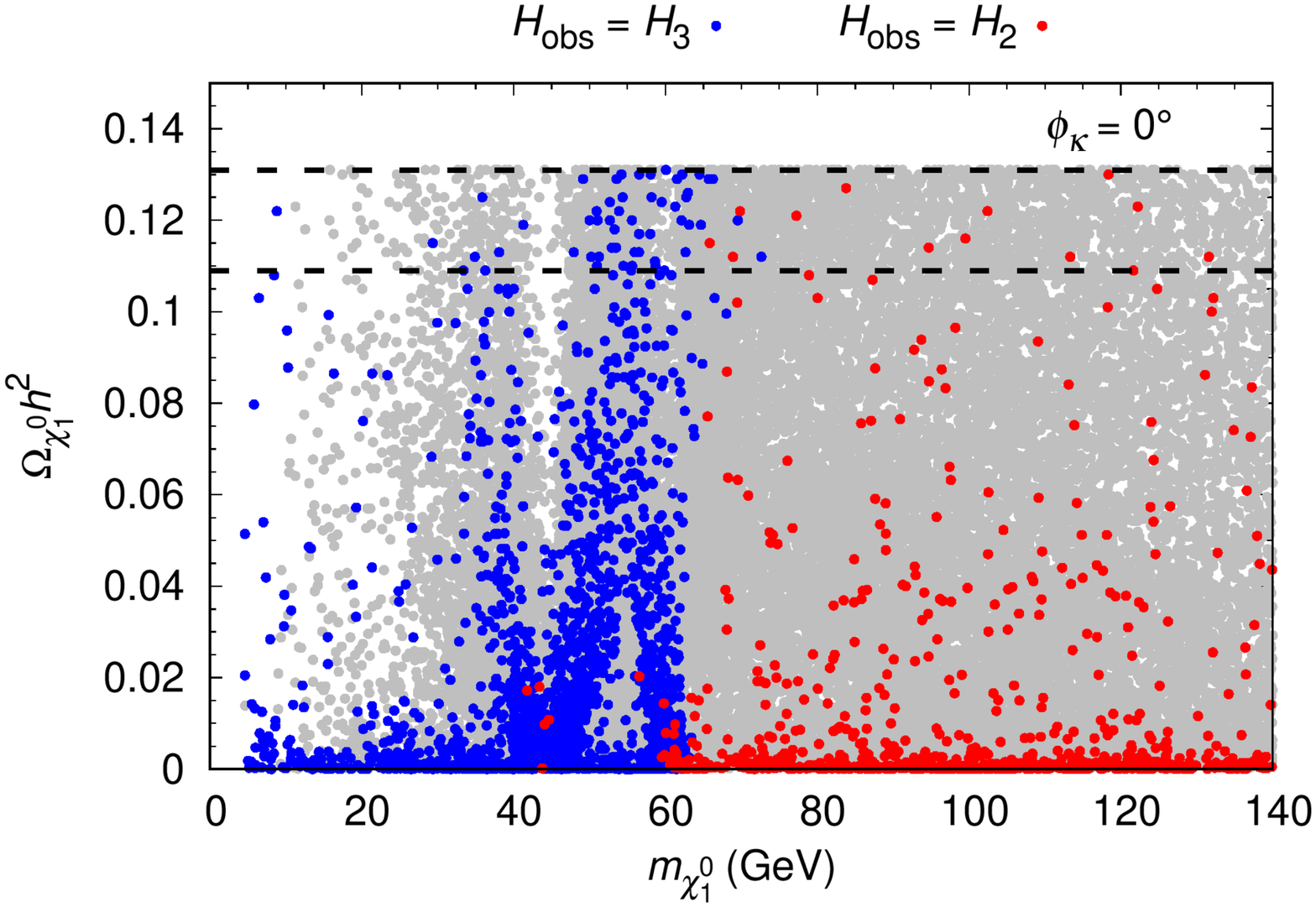} &
\hspace*{-1.2cm}\includegraphics*[width=9.5cm]{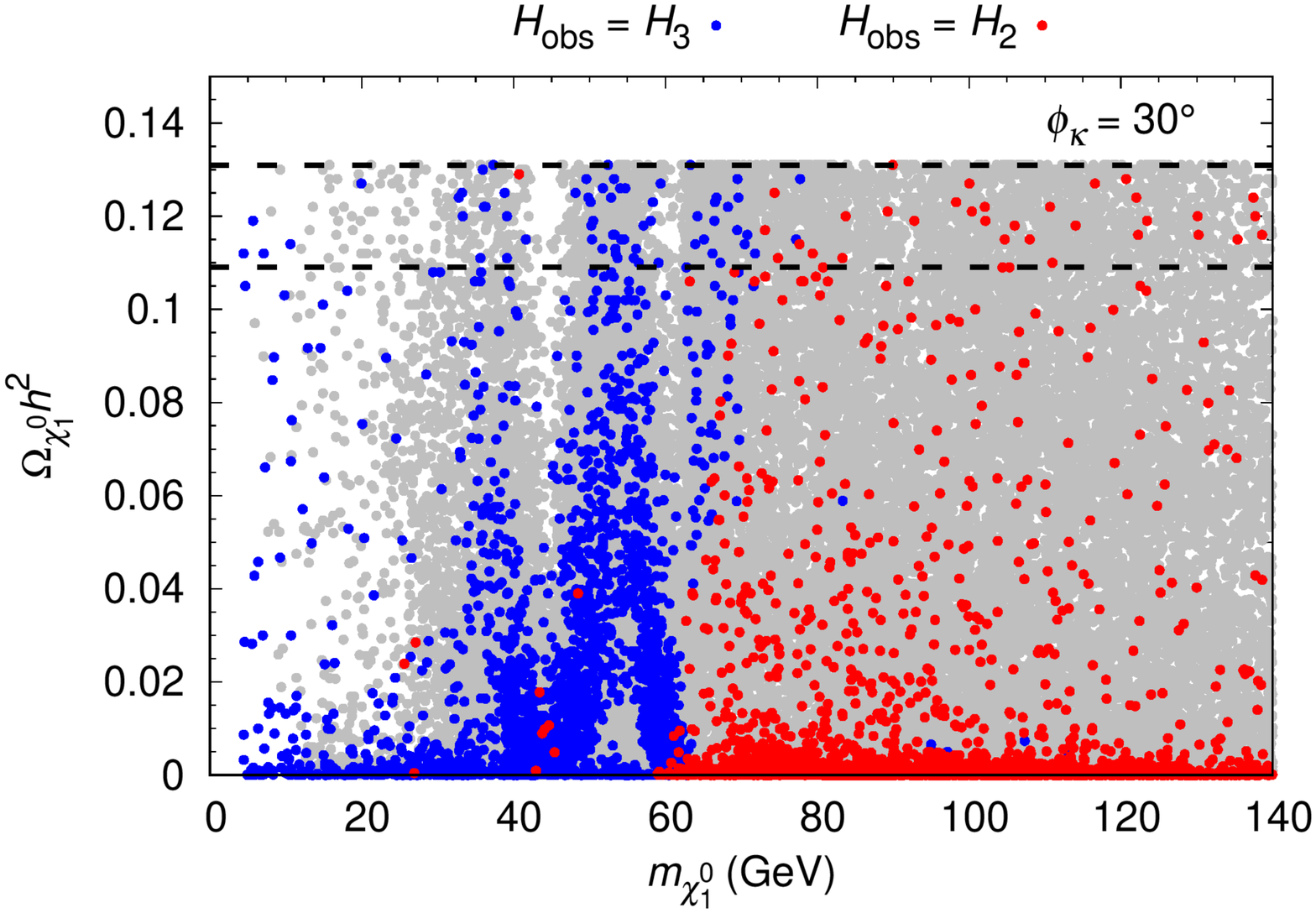} \\
\vspace*{-1.5cm}\hspace*{-0.9cm}\includegraphics*[width=9.5cm]{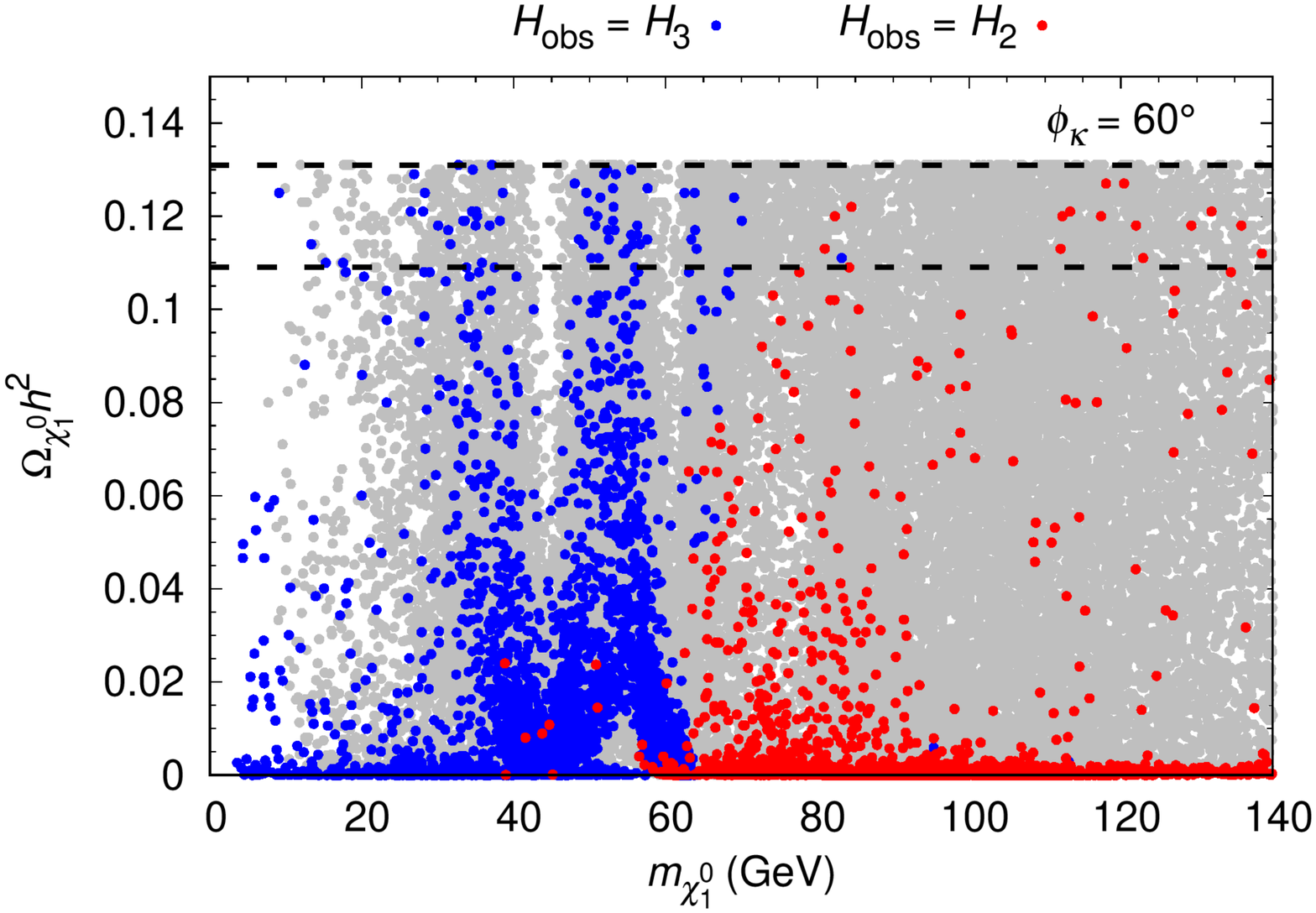} &
\hspace*{-1.2cm}\includegraphics*[width=9.5cm]{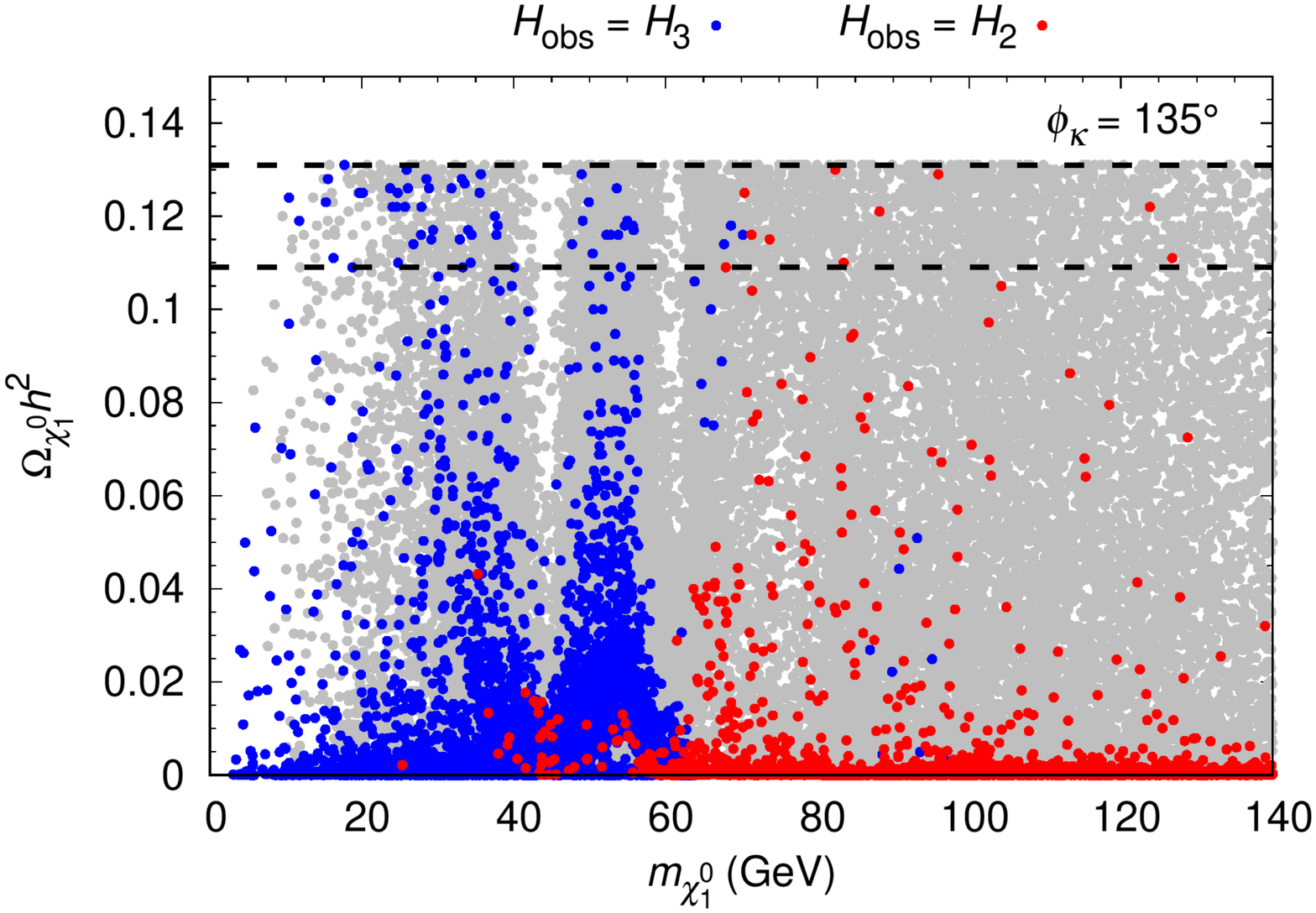}\\
\vspace*{-1.0cm}\hspace*{-0.9cm}\includegraphics*[width=9.5cm]{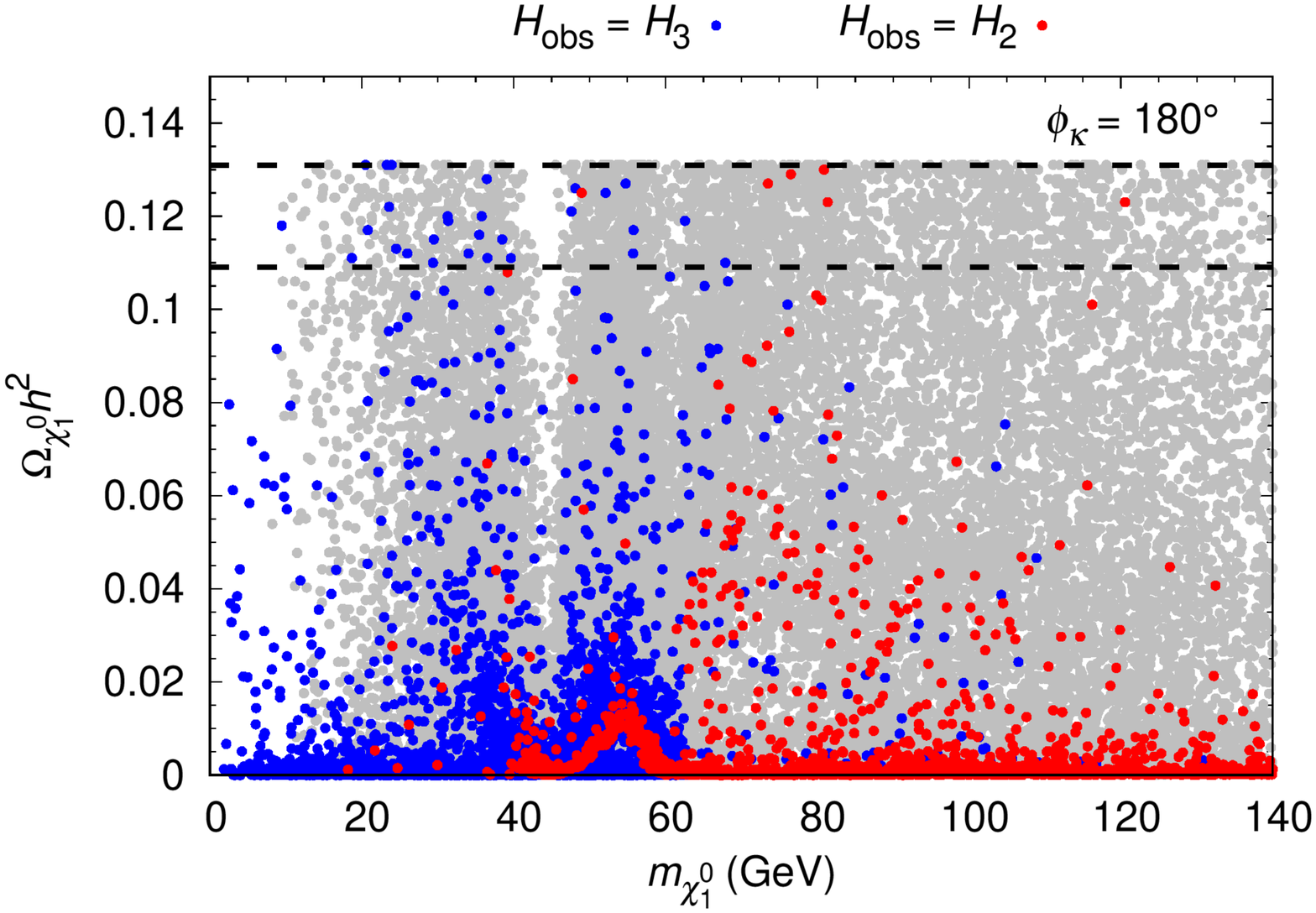} &
\hspace*{-1.2cm}\includegraphics*[width=9.5cm]{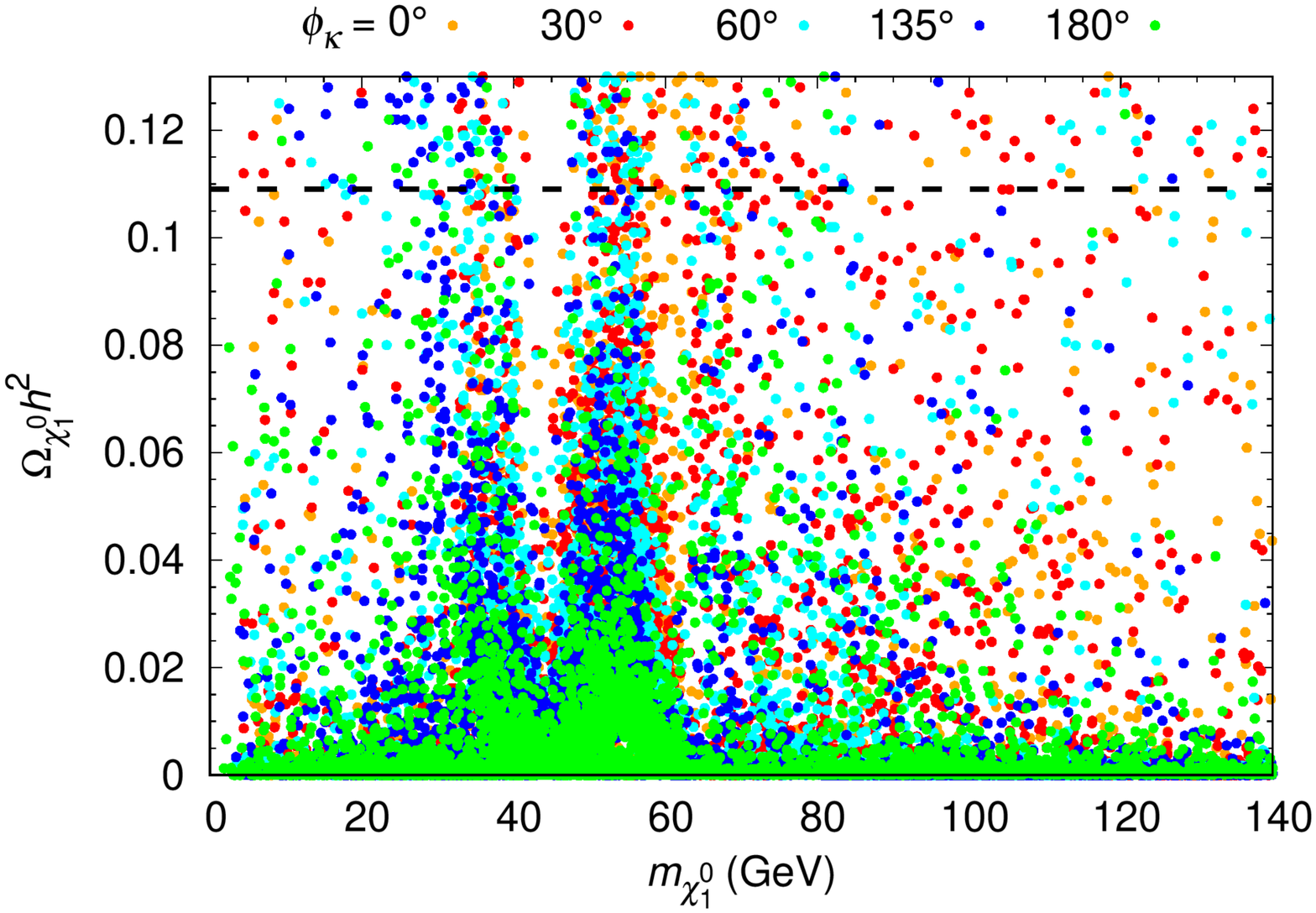} \\
\end{tabular}
\caption{\label{fig:0-180} Relic abundance of the DM as a function of its mass, for points obtained from the scans with $\phi_\kappa$ fixed to $0^\circ$ (top-left), $30^\circ$ (top-right), $60^\circ$ (centre-left), $135^\circ$ (centre-right), and $180^\circ$ (bottom-left). The red points correspond to the $\hsm=H_2$ scenario and the blue points to the $\hsm=H_3$ one. The grey points in the background are the ones ruled out by the PandaX-4T limits on $\sigma_{\rm SI}^p$. The bottom-right panel shows the allowed points from all the other panels overlapped.}
\end{figure}

In the top-left panel of the Fig.~\ref{fig:0-180}, corresponding to the CP-conserving case, $\Omega_{\neut{1}}{h^2}$ is generally quite small, except near $m_{\neut{1}}\sim m_Z/2$ and $m_{\neut{1}}\sim m_{H_{\rm SM}}/2$ for the $\hsm = H_3$ scenario (blue points), where a few points show consistency with the Planck measurement within $\pm 10\%$. A narrow peak of points also appears around $m_{\neut{1}}\sim 10$\,GeV, where, as we will see later, a very singlino-like $\neut{1}$ can undergo just the right amount of self-annihilation via the singlet $A_1$. Recall that all the five Higgs bosons are ordered by mass and not distinguished by their CP-assignment, and thus the $H_1$ in the cNMSSM can be either one of the $H_1$ or $A_1$ of the rNMSSM. In the $\hsm = H_2$ scenario (red points) the correct $\Omega_{\neut{1}}{h^2}$ can be obtained for a wide range of $m_{\neut{1}}$, when it is near either $m_{H_2}/2$ or, more frequently, $m_{H_3}/2$. For $\phi_\kappa = 30^\circ$, in the top-right panel, a few points with $m_{\neut{1}}$ between 10--20\,GeV also appear within (or just outside) the Planck band (i.e., $\Omega h^2 = 0.119\pm 10\%$). This is not the case for the CP-conserving case above, although the overall picture looks very similar, and is a result of the slight modification in the $\neut{1}$ composition owing to the CP-violating phase. 

When $\phi_\kappa$ is increased to $60^\circ$ (centre-left panel) some Planck-consistent points show up also around $m_{\neut{1}}= 30$\,GeV. Most notably, however, it is possible to obtain the correct $\Omega_{\neut{1}}{h^2}$ for the entire $\sim5-40$\,GeV mass range when the sign of $\kappa$ (and hence also of $\akap$) is flipped, as demonstrated by the centre-right and bottom-left panels corresponding to $\phi_\kappa = 135^\circ$ and $\phi_\kappa = 180^\circ$, respectively. In fact, for the latter phase, a sole point appears within the Planck band for $m_{\neut{1}}$ in the $\sim 40-45$\,GeV range (which is, however, excluded by the LHC electroweakino searches). This is not observed for any of the other selected phases in this figure, but we will discuss below a point for which it is achieved around $\phi_\kappa = 60^\circ$ also. The bottom-right panel  presents a holistic picture, where one sees that nearly the entire sub-100\,GeV range of $m_{\neut{1}}$ with $\Omega h^2 = 0.119\pm 10\%$ is covered by the good points from all of the scans.  

For a closer analysis of the impact of the variation in $\phi_\kappa$ on $m_{\neut{1}}$ and its relic abundance, we selected four test points (TPs) from among those corresponding to $\phi_\kappa = 30^\circ$. The values of the corresponding scanned parameters, along with the spectra for $\phi_\kappa= 30^\circ$, are given in Table \ref{TAB:benchmarks}. For each of these TPs, $m_{\neut{1}}$ is plotted as a function of $\phi_\kappa$ in all the panels of Fig. \ref{fig:BPs-masses}. The heat map in the left column of the figure corresponds to $m_{H_1}$ and in the right column to $m_{H_2}$. In this as well as the two figures that follow, the non-existence of a point for some values of $\phi_\kappa$ in a given panel implies that {\tt SPheno} did not produce an output on account of there being unphysical loop-corrected masses for some particles. The grey points imply inconsistency with one (or more) of the constraints 1--5 listed in section \ref{sec:numeric} and the two additional limits from the LHC noted above, while the coloured circles give $\Omega_{\neut{1}}{h^2}>0.131$. The coloured boxes instead mean $\Omega_{\neut{1}}{h^2}<0.131$ for that point, and a cross around a box reflects that $\Omega_{\neut{1}}{h^2}$ lies within the Planck band. 
 
The TP1 in the top row of Fig.~\ref{fig:BPs-masses} is the single (red) point for the $\hsm=H_2$ scenario with $m_{\neut{1}}\sim 40$\,GeV appearing within the Planck band for $\phi_\kappa = 30^\circ$. It does so, however, only for this specific value of $\phi_\kappa$. For almost the entire remaining range of the phase, this parameter space configuration is inconsistent with at least one of the enforced experimental constraints. 
The remaining three TPs belong to the $\hsm=H_3$ scenario. For TP2 also, the Planck-consistent amount of self-annihilation of the $\neut{1}$, via the $H_2$, occurs only for $\phi_\kappa$ a few degrees around $30^\circ$. $m_{\neut{1}}$ and $m_{H_2}$ both reduce with increasing $\phi_\kappa$ - the latter much slower than the former - until tachyonic masses appear in the particle spectrum for $\phi_\kappa \geq 80^\circ$. In the case of TP3, as with the TP1, $\Omega_{\neut{1}} h^2 = 0.119\pm 10\%$ is satisfied only for $\phi_\kappa = 30^\circ$, when the sharply falling $m_{H_1}$ gets very close to $2 m_{\neut{1}}\simeq 12$\,GeV, as seen in the third row of the left panel. Beyond this value of $\phi_\kappa$, the $\Omega_{\neut{1}} h^2$ drops for a few degrees, owing to excessive annihilation, before rising above the Planck bound again when $m_{H_1}$ grows too small.
Finally, TP4 is a representative point of the case when the $\Omega_{\neut{1}} h^2$ falls within the Planck band for $m_{\neut{1}}$ in the $\sim 40-45$\,GeV range, as hinted earlier. While this TP has also been taken from among the good points for $\phi_\kappa=30^\circ$ in Fig.~\ref{fig:0-180}, its $\Omega_{\neut{1}} h^2$ lies below the Planck band for the original $\phi_\kappa$. 

\begin{figure}[tbp]
\begin{tabular}{cc}
\vspace*{-2.2cm}\hspace*{-1.0cm}\includegraphics*[width=9.5cm]{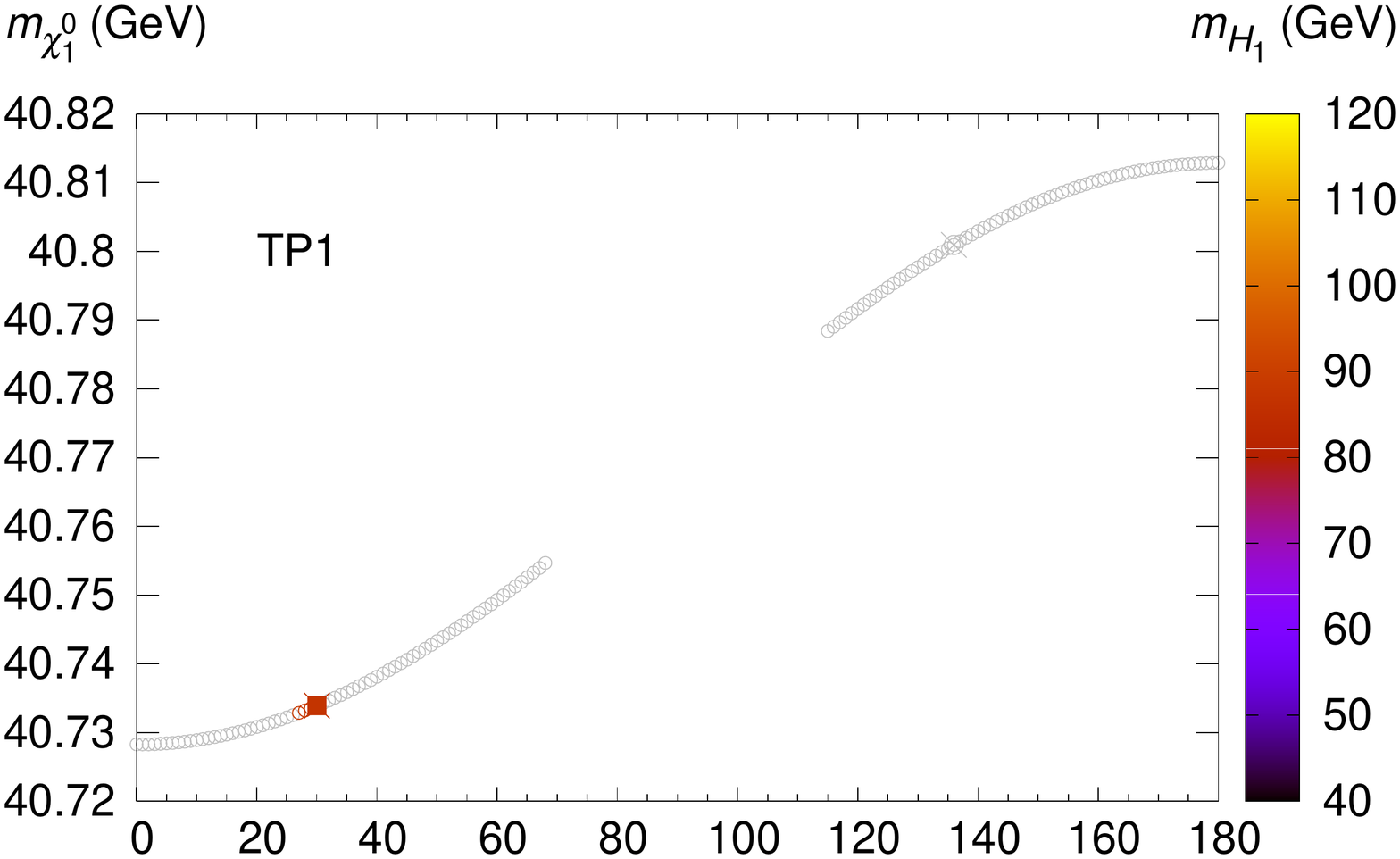} &
\hspace*{-1.3cm}\includegraphics*[width=9.5cm]{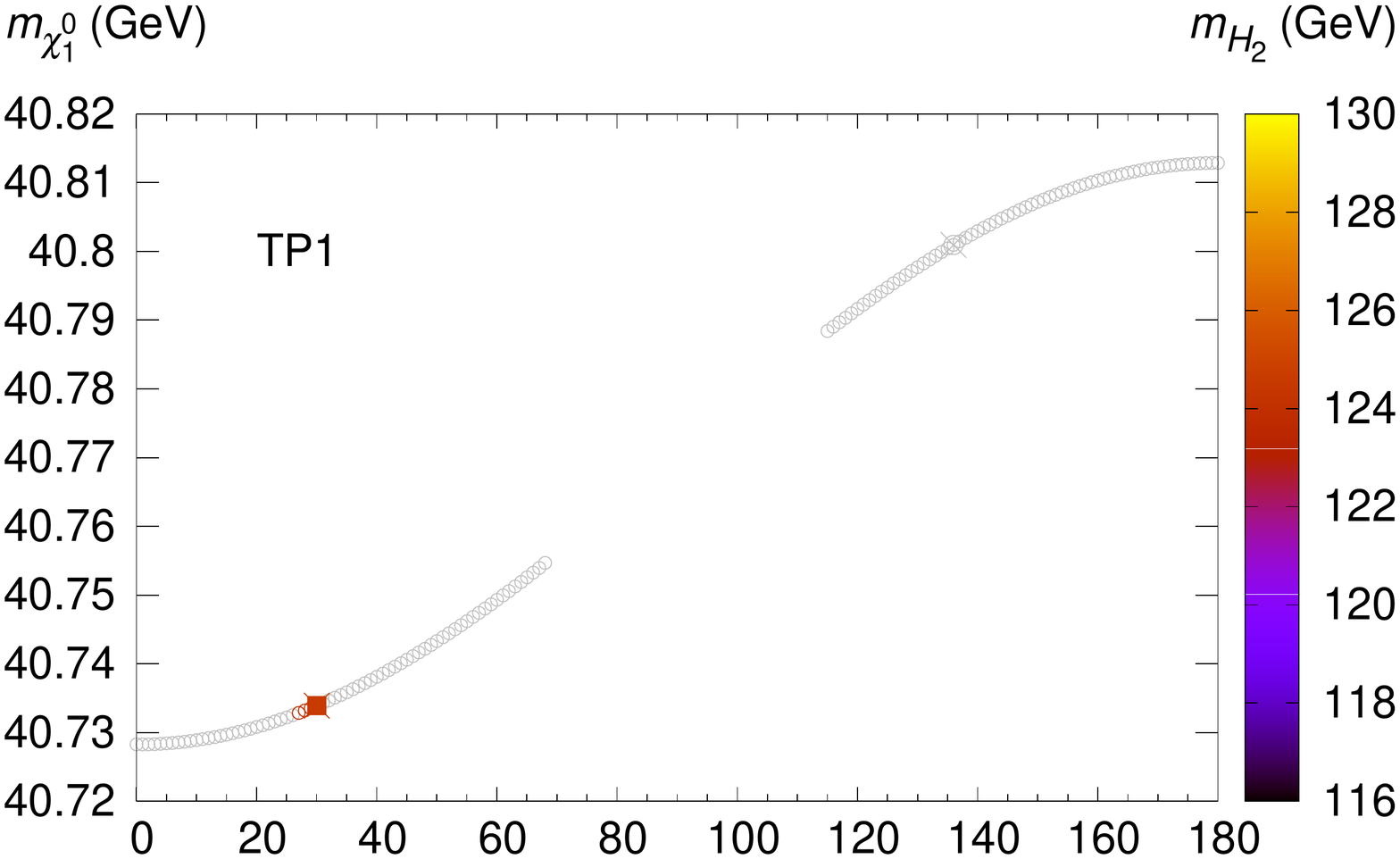} \\
\vspace*{-2.2cm}\hspace*{-1.0cm}\includegraphics*[width=9.5cm]{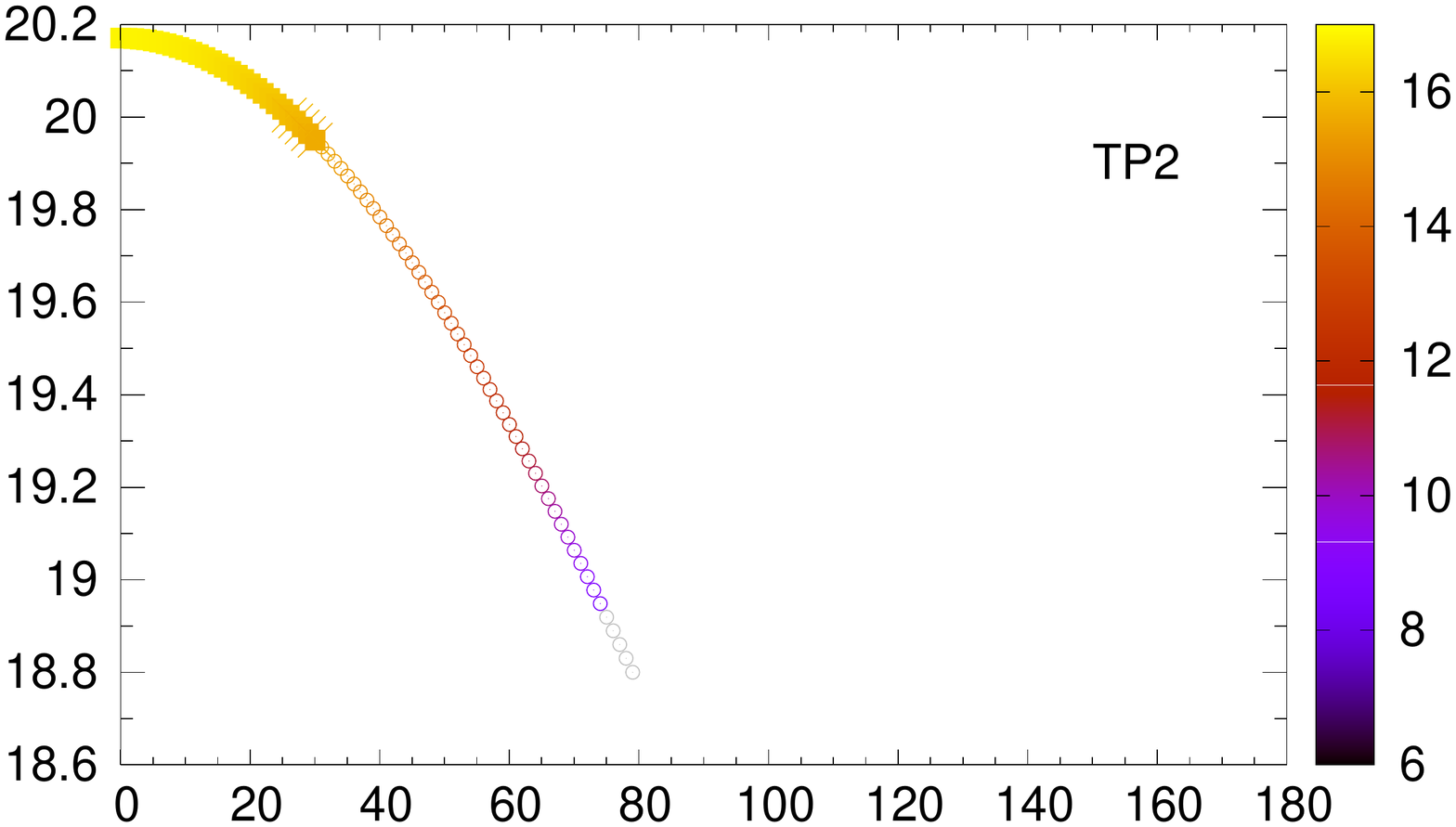} &
\hspace*{-1.3cm}\includegraphics*[width=9.5cm]{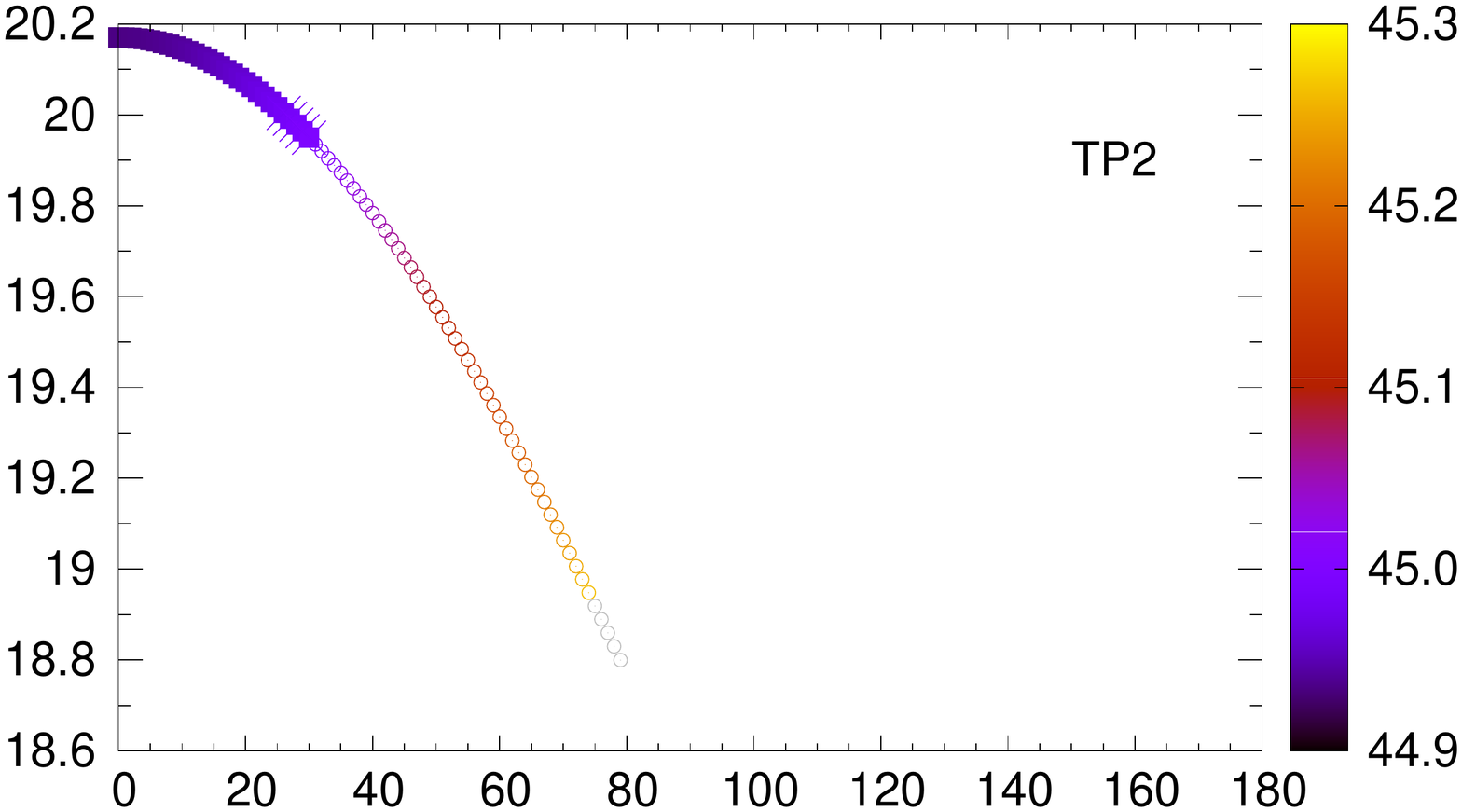} \\
\vspace*{-2.2cm}\hspace*{-1.0cm}\includegraphics*[width=9.5cm]{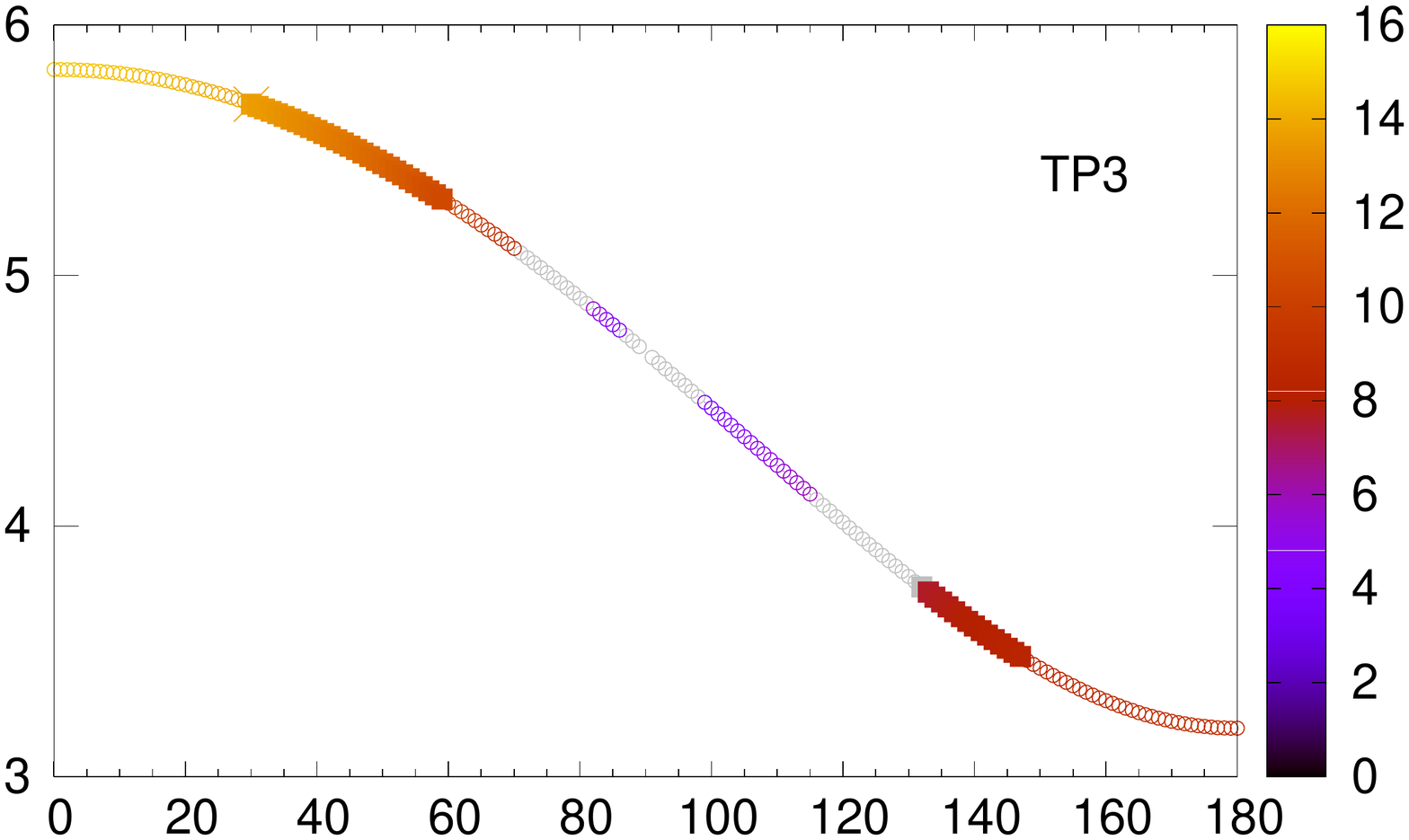} &
\hspace*{-1.3cm}\includegraphics*[width=9.5cm]{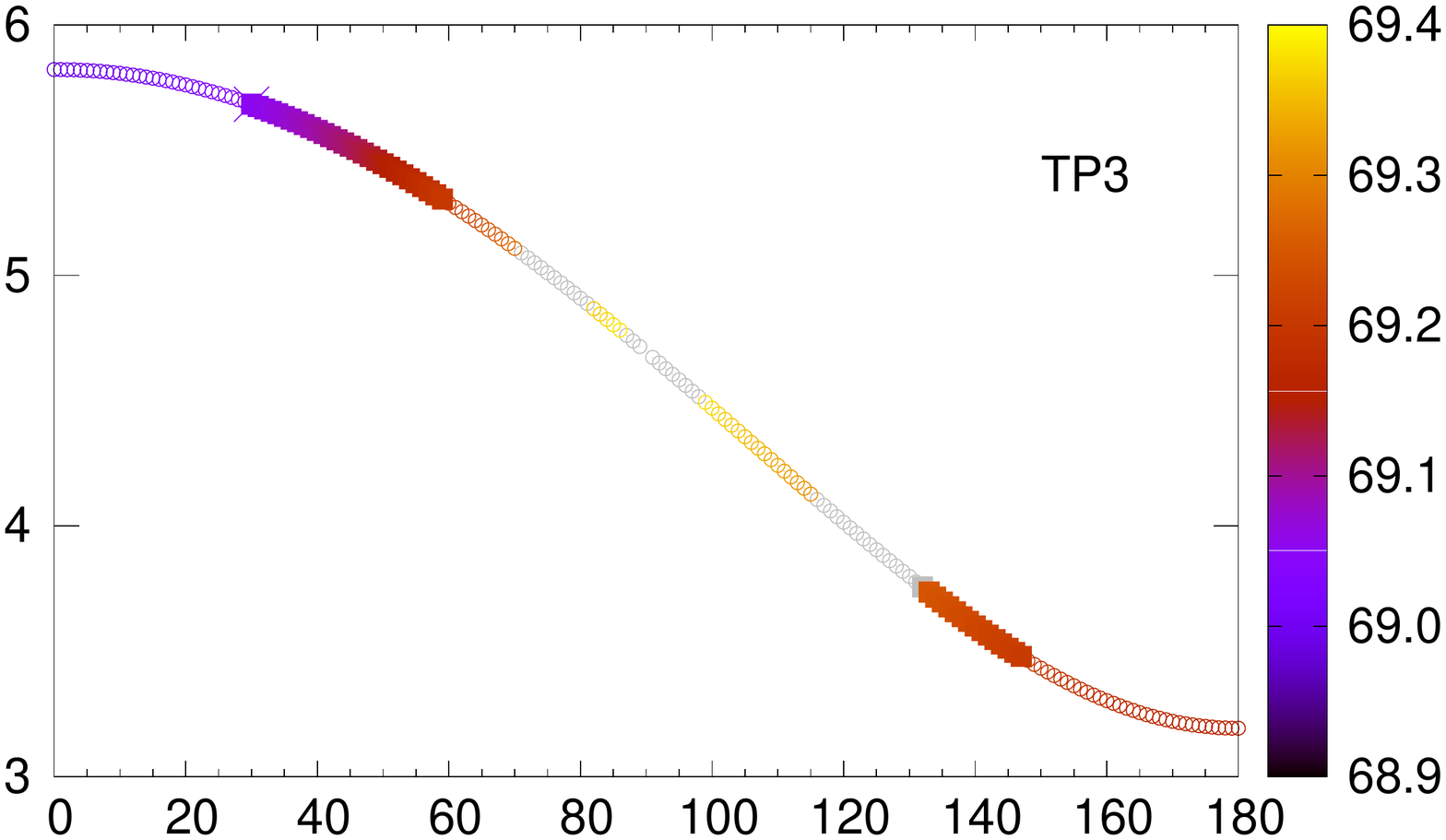} \\
\vspace*{-1.0cm}\hspace*{-1.0cm}\includegraphics*[width=9.5cm]{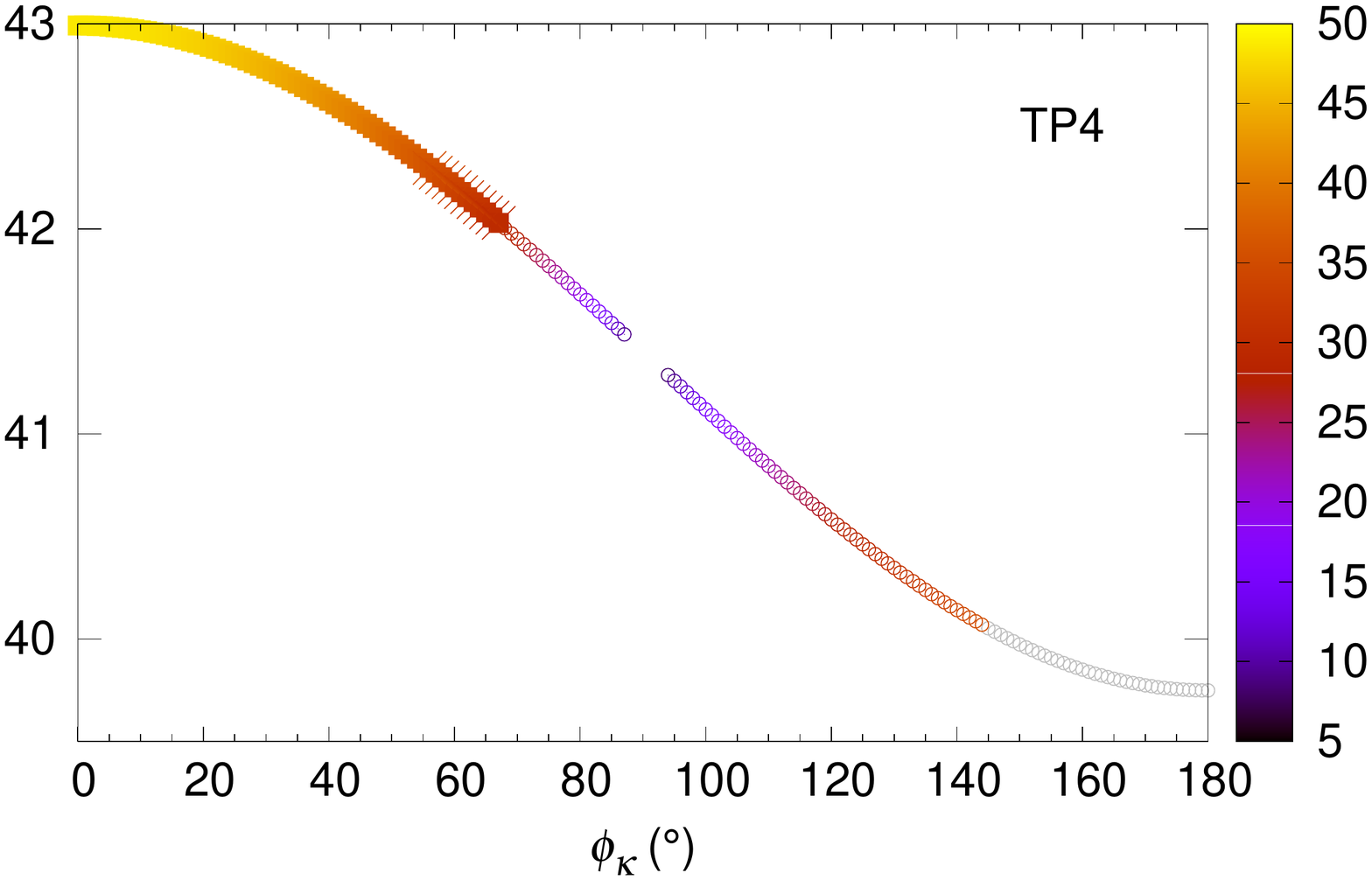} &
\hspace*{-1.3cm}\includegraphics*[width=9.5cm]{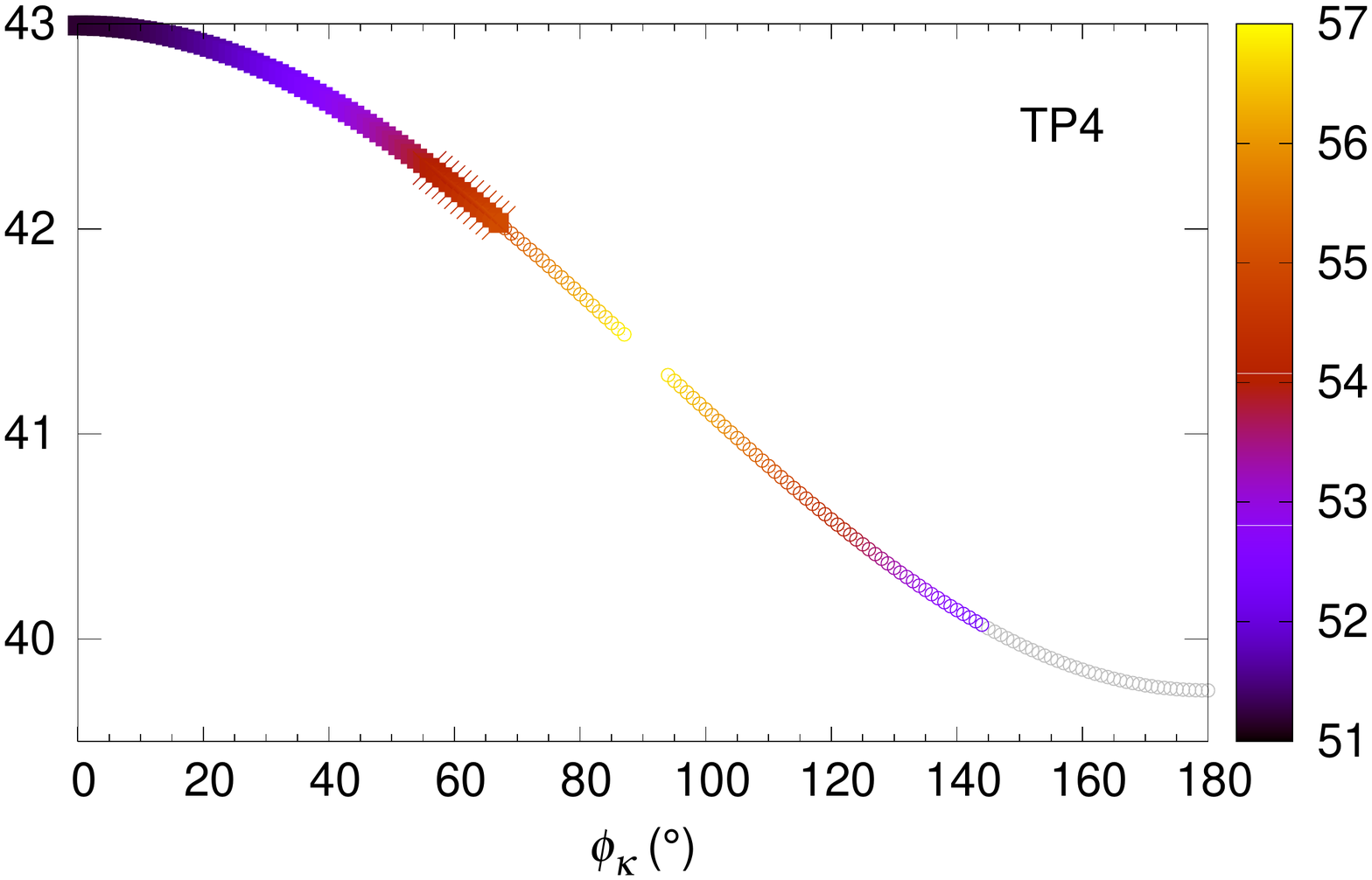} \\
\end{tabular}
\caption{\label{fig:BPs-masses} $m_{\neut{1}}$ as a function of $\phi_\kappa$ for the test points 1 (top row) -- 4 (bottom row). The heat maps correspond to $m_{H_1}$ (left column), $m_{H_2}$ (right column). The grey circles imply inconsistency with one of the experimental constraints, while the coloured circles give $\Omega_{\neut{1}}{h^2}>0.131$. The coloured boxes correspond to $\Omega_{\neut{1}}{h^2}<0.131$, and a cross around a box implies $0.107 < \Omega_{\neut{1}}{h^2}<0.131$, besides consistency with all the other constraints.}
\end{figure}

In the left column of Fig. \ref{fig:BPs-mass-BR} $m_{\neut{1}}$ is again plotted as a function of $\phi_\kappa$ for the TPs 1--4 (top row to bottom row), with the heat map now depicting the singlino fraction, $N_s$, of the $\neut{1}$. For TP1, the $\neut{1}$ has a negligible singlino component, but is instead entirely bino-like, with a small higgsino fraction just enough for the correct amount of its self-annihilation via the $Z$ boson for $\phi_\kappa = 30^\circ$. On the other hand, the very large $N_s$ in the CP-conserving case for TP2 falls sharply with increasing $\phi_\kappa$, with the Planck-consistency occurring when it is just above 90\% around $\phi_\kappa \sim 30^\circ$. For TP3 the $N_s$ stays almost constant over the entire range of $\phi_\kappa$, while for TP4, as the singlino fraction as well as the mass of $\neut{1}$ drop slowly with increasing $\phi_\kappa$, its $Z$-mediated annihilation gradually reduces. It reaches a level sufficient to give the correct $\Omega_{\neut{1}} h^2$ for $\phi_\kappa \sim 55^\circ -65^\circ$. The right column of this figure shows the BR($\hsm \to \neut{1}\neut{1}$). For TP1 it fluctuates between 1.5\% and 2\% for the allowed values of $\phi_\kappa$, and for TP2 it rises with $\phi_\kappa$ but does not exceed 4\%. For TP3 the BR$(\hsm\to \neut{1}\neut{1})$ rises noticeably with $\phi_\kappa$ (while $m_{\neut{1}}$, given by the heat map, drops), until it reaches the maximum of about 10\% for 180$^\circ$, while for TP4 it is always insignificant.

In Fig.\ref{fig:BPs-BRs} we take a brief look at the BRs of the $\hsm$ into $H_1H_1$ (left column) and $H_2H_2$ (right column), in a bid to further understand the implications of different $\phi_\kappa$ for the $\hsm$ phenomenology at the LHC. For TP1, the BR$(H_2\to H_1H_1)$ is vanishing for the Planck-consistent $\phi_\kappa=30^\circ$, while the $H_3 \to H_2 H_2$ decay is kinematically forbidden. In the case of TP2, the BR$(H_3\to H_1H_1)$ drops from about 12\% in the CP-conserving case to about 8\% for $\phi_\kappa = 80^\circ$, but the BR$(\hsm\to H_2H_2)$ increases by about 2\%. The BR$(H_3\to H_1H_1)$ for TP3 also falls by about 4\% overall, as in the case of TP2. For TP4, the BR$(H_3\to H_1H_1)$ drops from about 7\% near $\phi_\kappa=0^\circ$ to less than 1\% for $\phi_\kappa = 180^\circ$, implying that the already slim prospects of observing the 4-body final state resulting from $\hsm \to H_1 H_1$ decay for this point further reduce significantly as the amount of CP-violation increases. At the same time, though, the BR$(\hsm\to H_2H_2)$ rises from $\sim$6\% to about 17\%, but this causes exclusion of $\phi_\kappa > 144^\circ$ by the LHC data. Overall then, the phenomenology of the 4-body final states with invariant mass near that of $\hsm = H_3$ for points analogous to the TP4 could be crucial for distinguishing the signatures of CP-conserving versus the CP-violating NMSSM. It will be the subject of a follow-up anaylsis.

\begin{figure}[tbp]
\begin{tabular}{cc}
\vspace*{-2.2cm}\hspace*{-1.0cm}\includegraphics*[width=9.5cm]{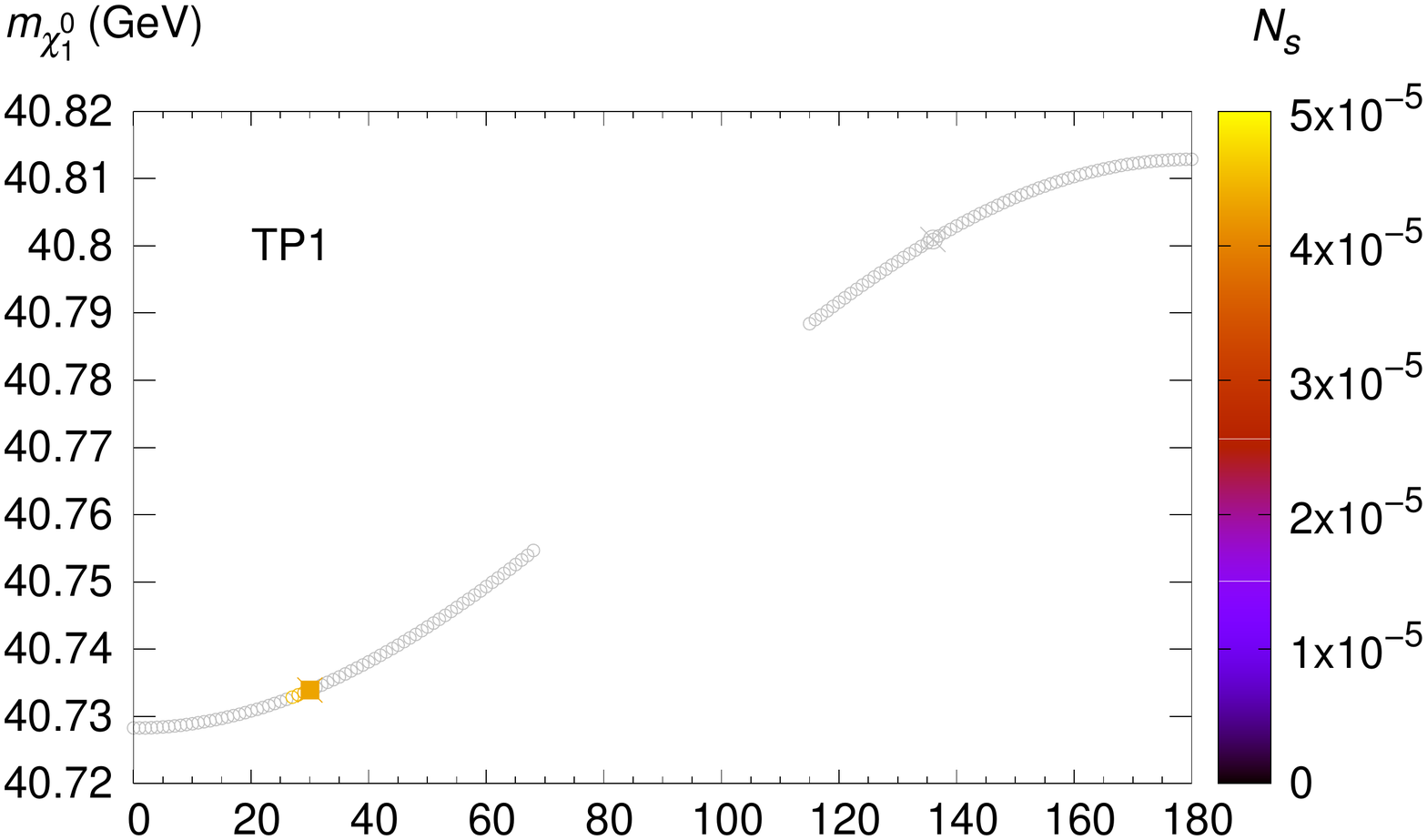} &
\hspace*{-1.5cm}\includegraphics*[width=9.5cm]{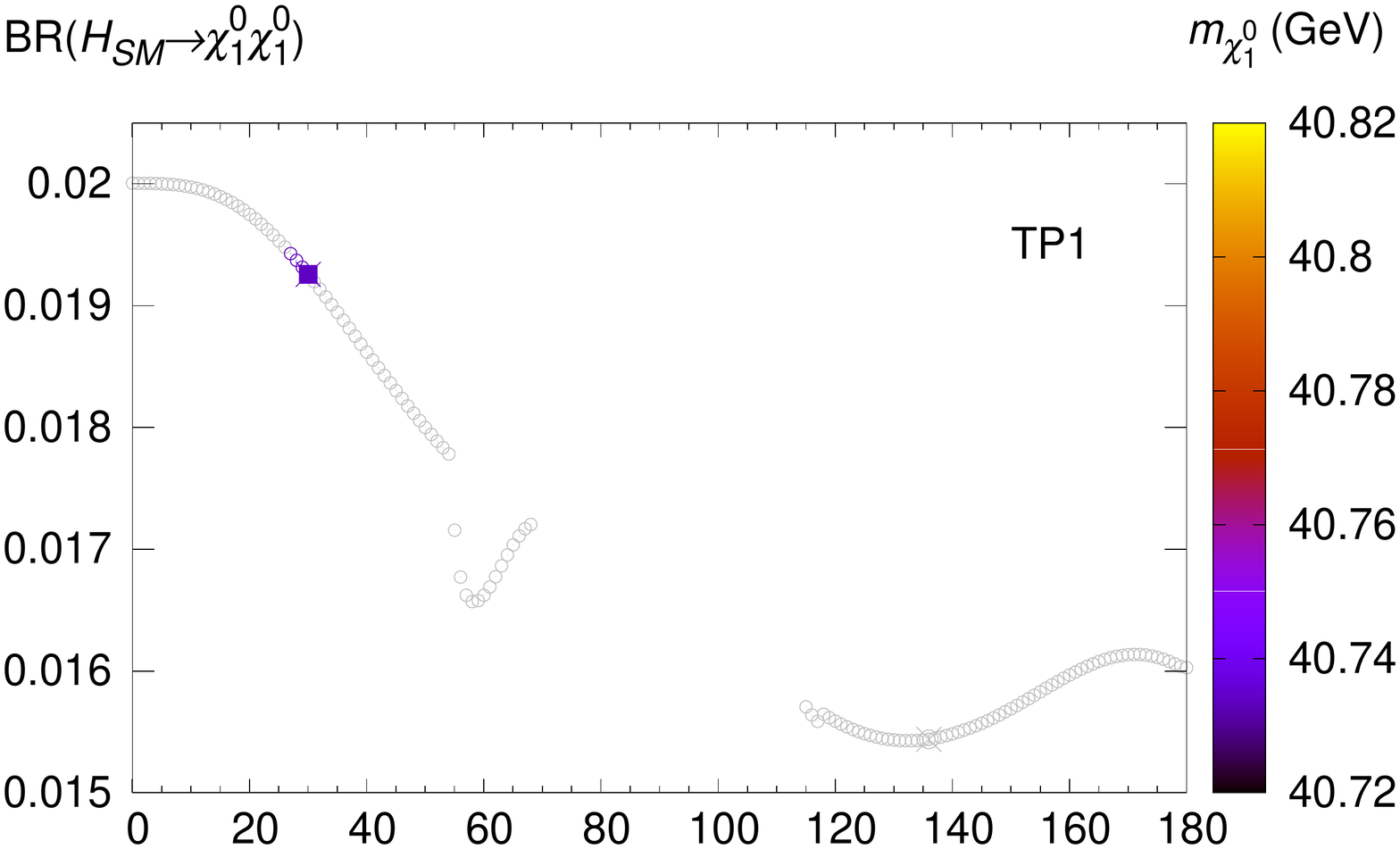} \\
\vspace*{-2.2cm}\hspace*{-1.0cm}\includegraphics*[width=9.5cm]{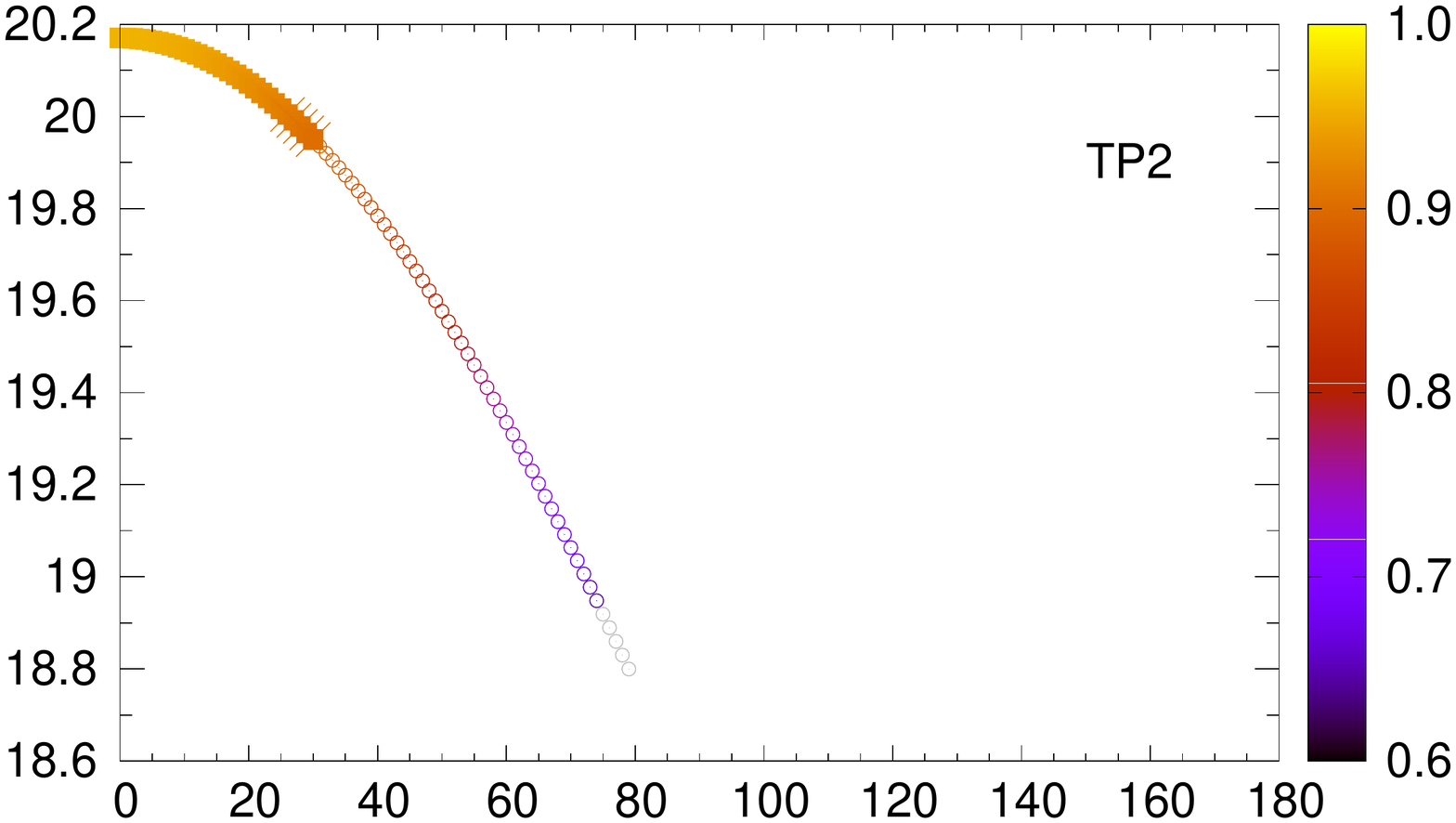} &
\hspace*{-1.5cm}\includegraphics*[width=9.5cm]{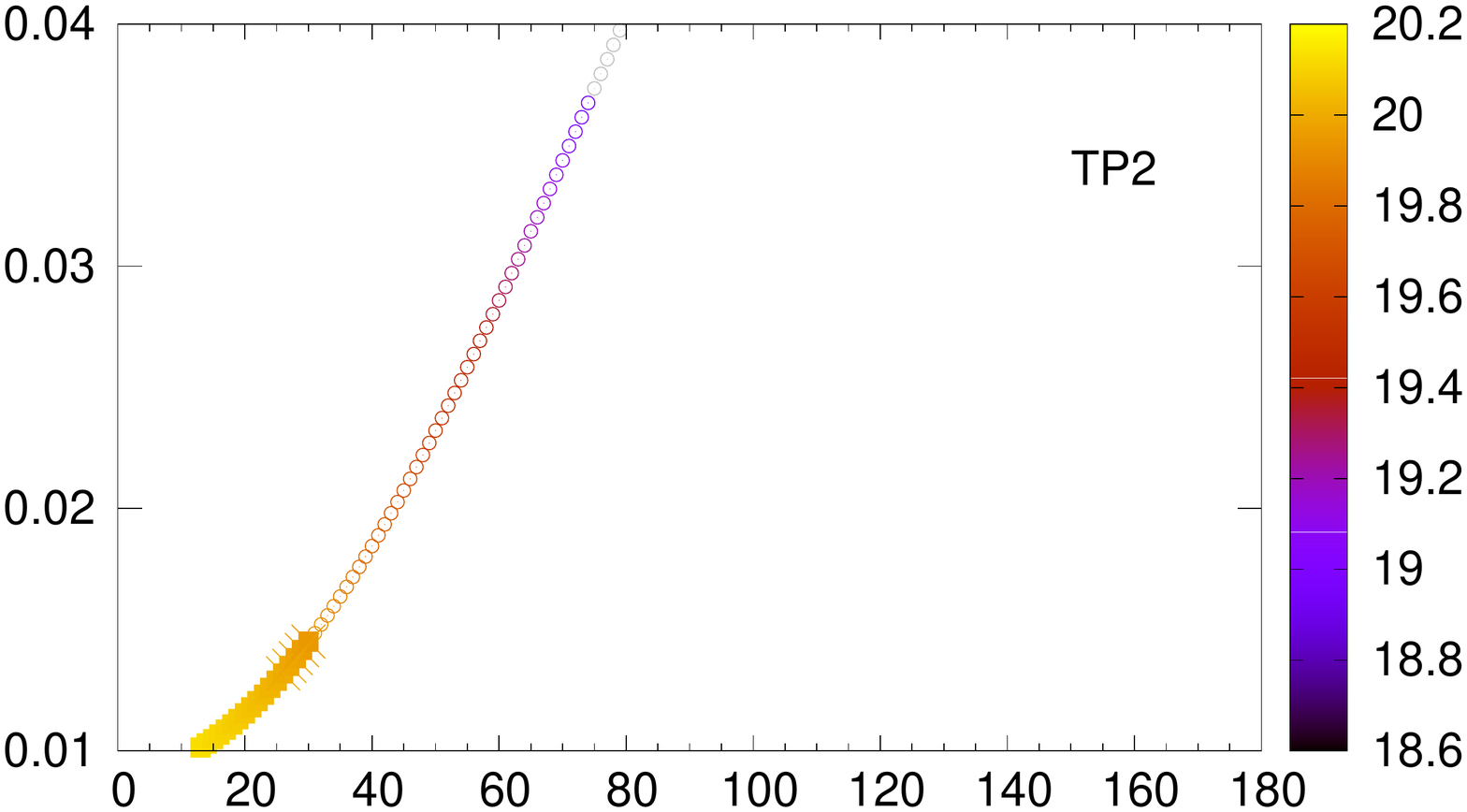} \\
\vspace*{-2.2cm}\hspace*{-1.0cm}\includegraphics*[width=9.5cm]{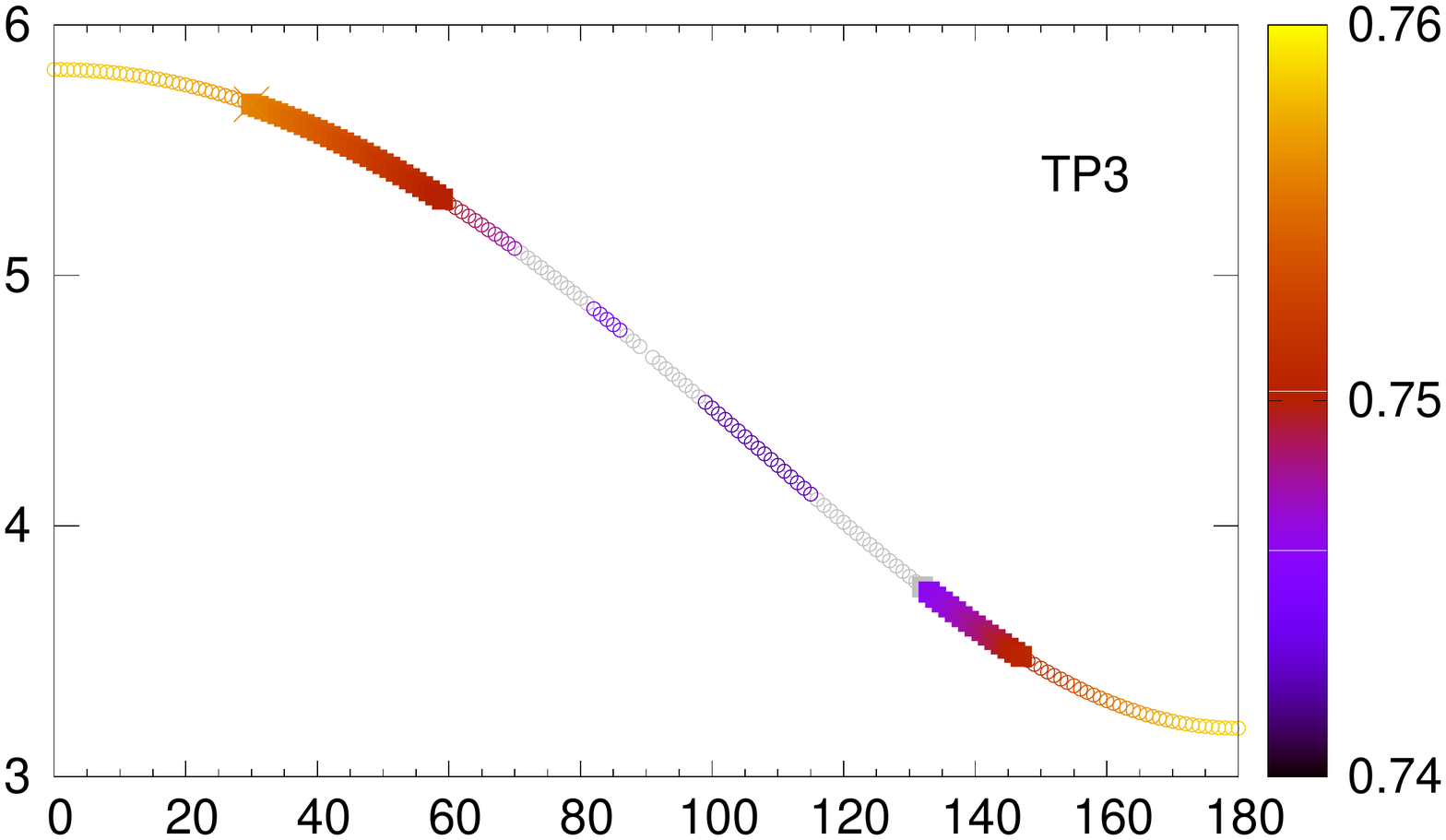} &
\hspace*{-1.5cm}\includegraphics*[width=9.5cm]{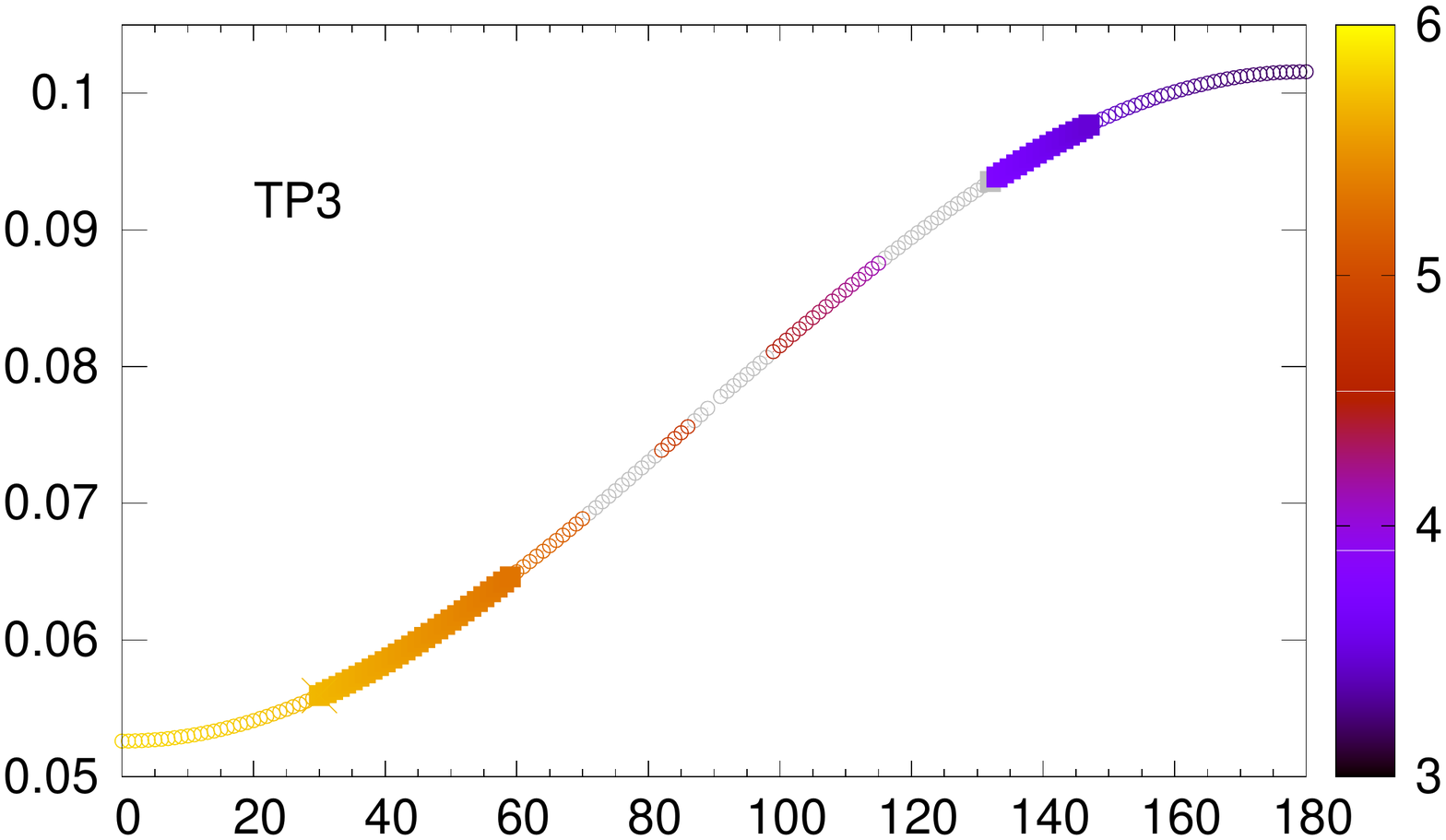} \\
\vspace*{-1.0cm}\hspace*{-1.0cm}\includegraphics*[width=9.5cm]{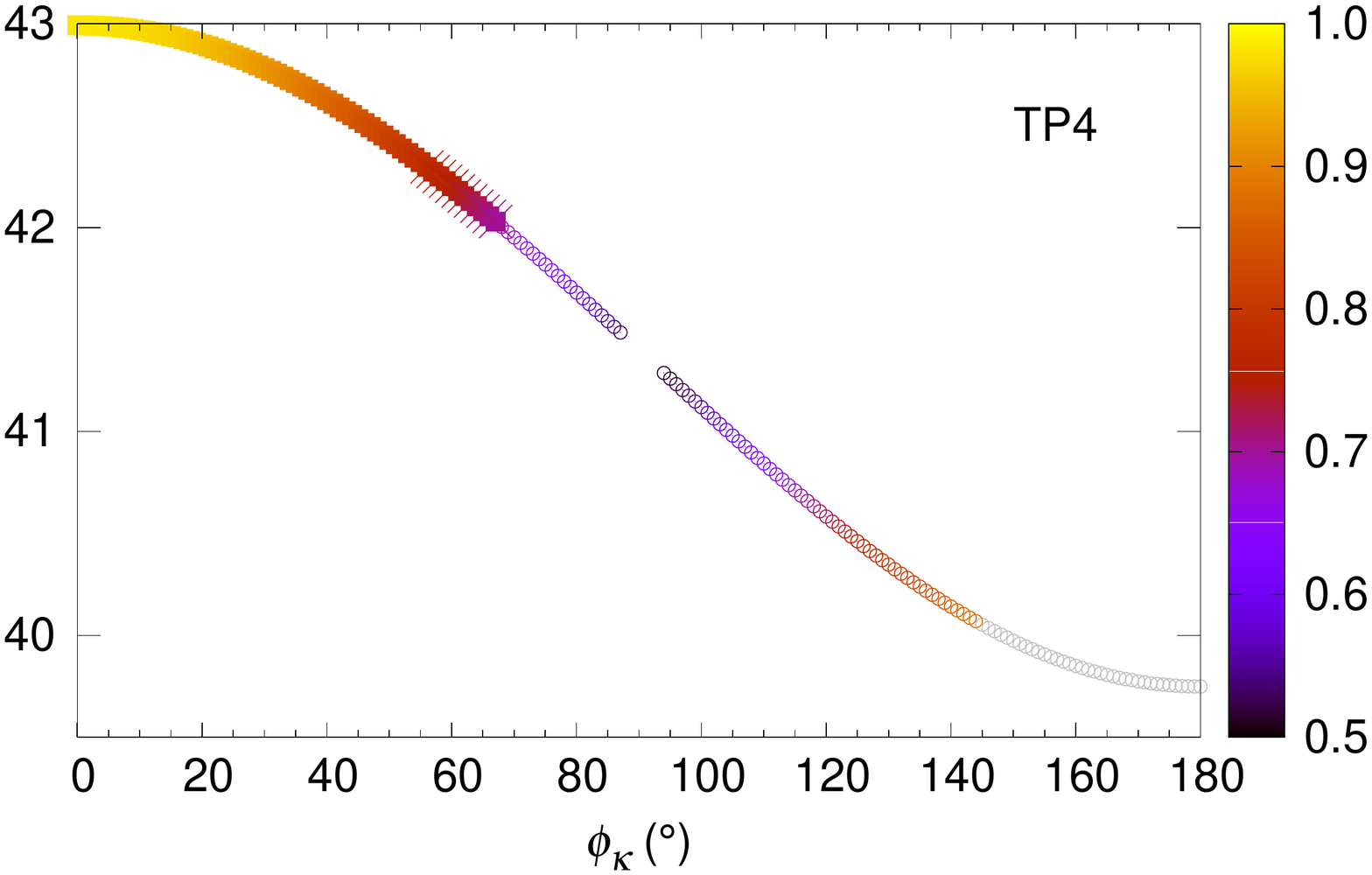} &
\hspace*{-1.5cm}\includegraphics*[width=9.5cm]{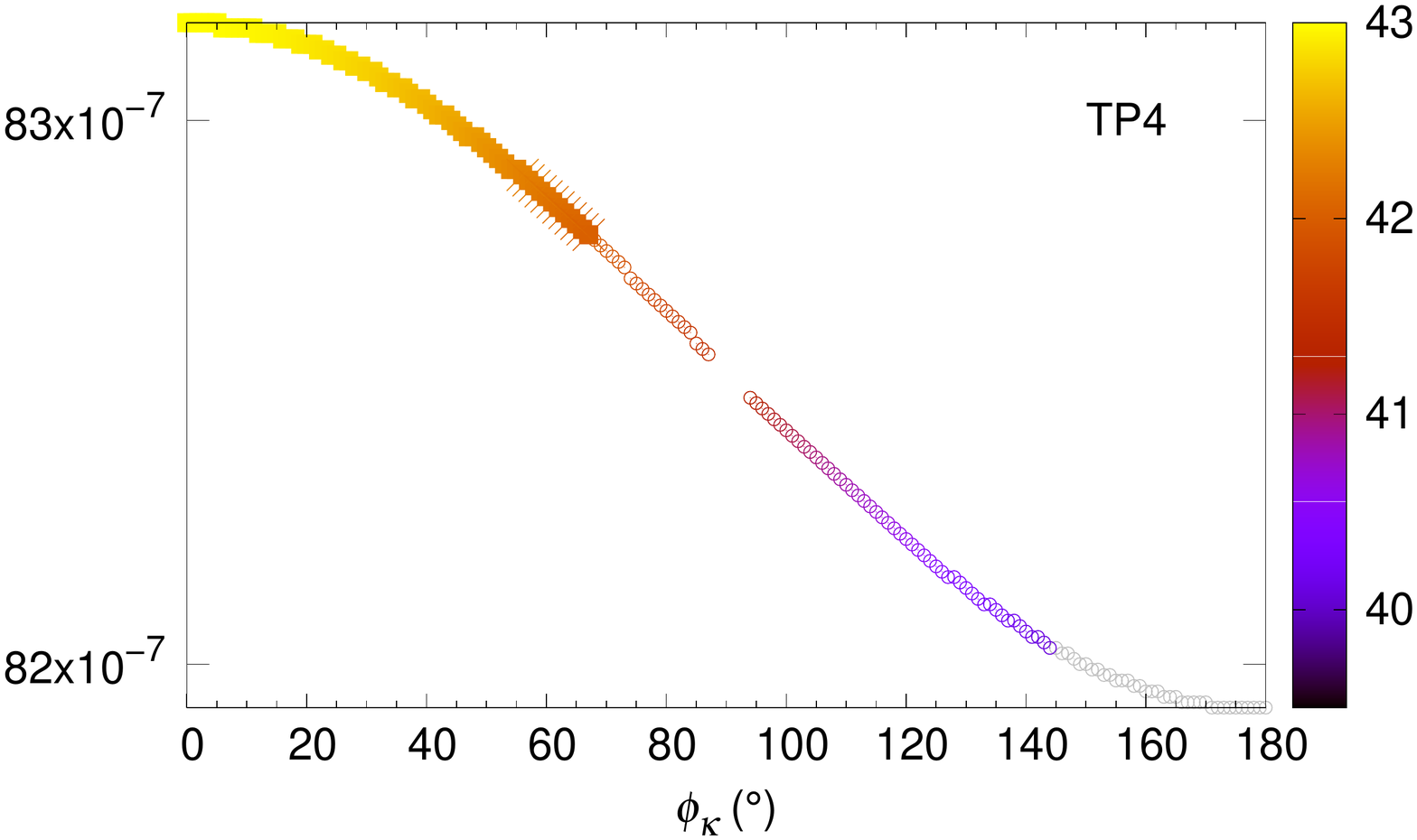} \\
\end{tabular}
\caption{\label{fig:BPs-mass-BR} $m_{\neut{1}}$ (left column) and BR($\hsm \to \neut{1}\neut{1}$) (right column) as functions of $\phi_\kappa$ for the test points 1 (top row) -- 4 (bottom row). The heat maps in the left and right columns correspond to $N_s$ and $m_{\neut{1}}$, respectively. The colouring scheme is the same as in Fig.~\ref{fig:BPs-masses}.}
\end{figure}

Finally, since the $\widetilde{\chi}^\pm_1$/$\neut{2}$ in all our TPs are higgsino-like and always heavier than 210\,GeV, they are consistent with the current exclusion limits from the LHC. Besides, instead of decaying to $W/Z/\hobs$, our $\neut{2}$ can decay dominantly to the lighter singlet-like Higgs boson(s) and thus keep evading detection in the near future. Likewise, the $\widetilde{\chi}^\pm_2$/$\neut{5}$ are wino-like and heavier than 750\,GeV for these four points. Nevertheless, in our follow-up analysis it would be interesting to test our scan points against latest results using a fast tool such as {\tt SModelS} \cite{Kraml:2013mwa,Ambrogi:2017neo,Ambrogi:2018ujg,Khosa:2020zar,Alguero:2021dig}, and to process a handful using full recasting in {\tt MadAnalysis} with the latest searches in \cite{Goodsell:2020ddr,Fuks:2021zbm} (see also \cite{LHCReinterpretationForum:2020xtr} for a review of available recasting tools), as performed in, e.g., \cite{Goodsell:2020lpx}.

\begin{figure}[tbp]
\begin{tabular}{cc}
\vspace*{-2.2cm}\hspace*{-1.0cm}\includegraphics*[width=9.5cm]{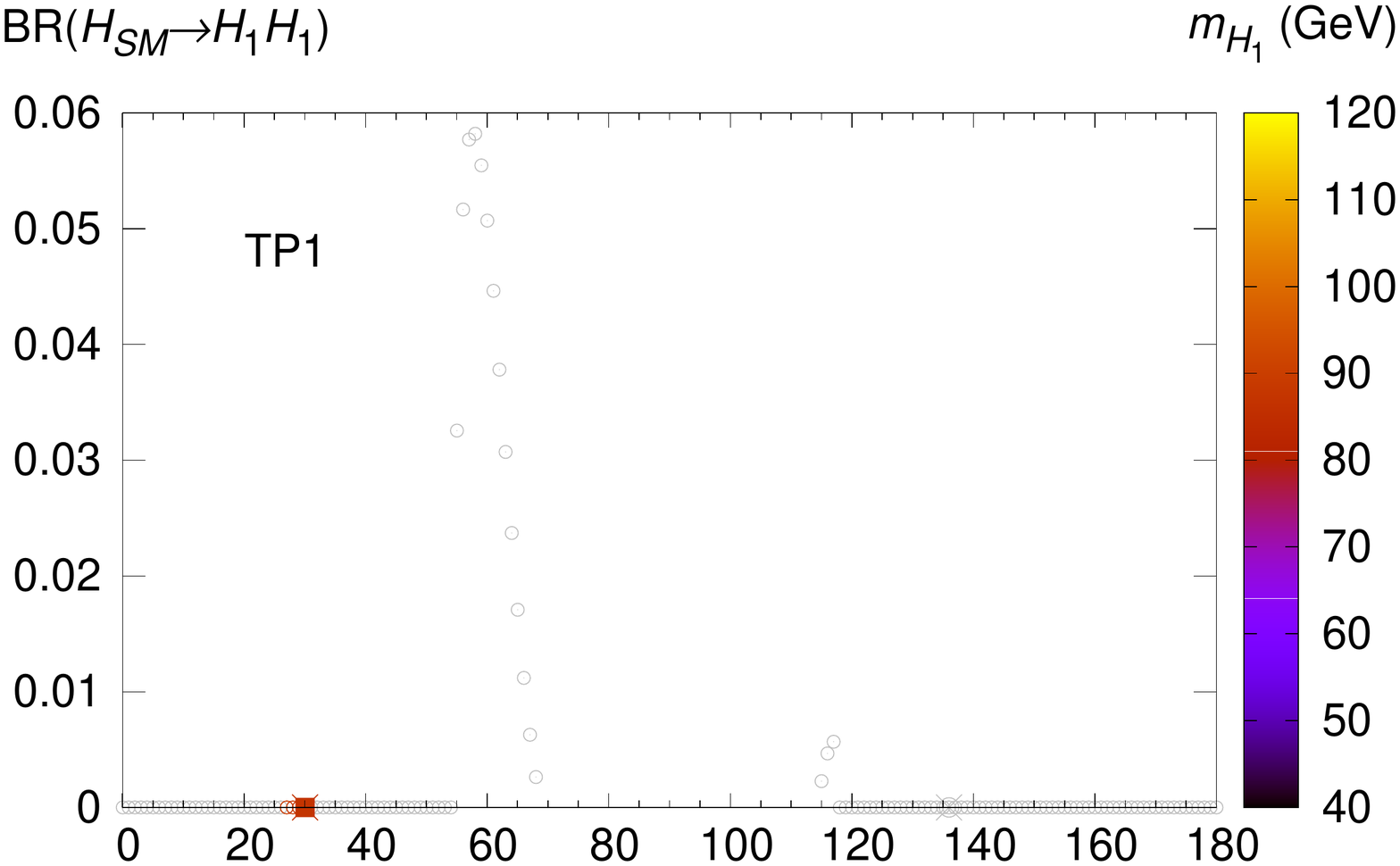} & \\
\vspace*{-2.2cm}\hspace*{-1.0cm}\includegraphics*[width=9.5cm]{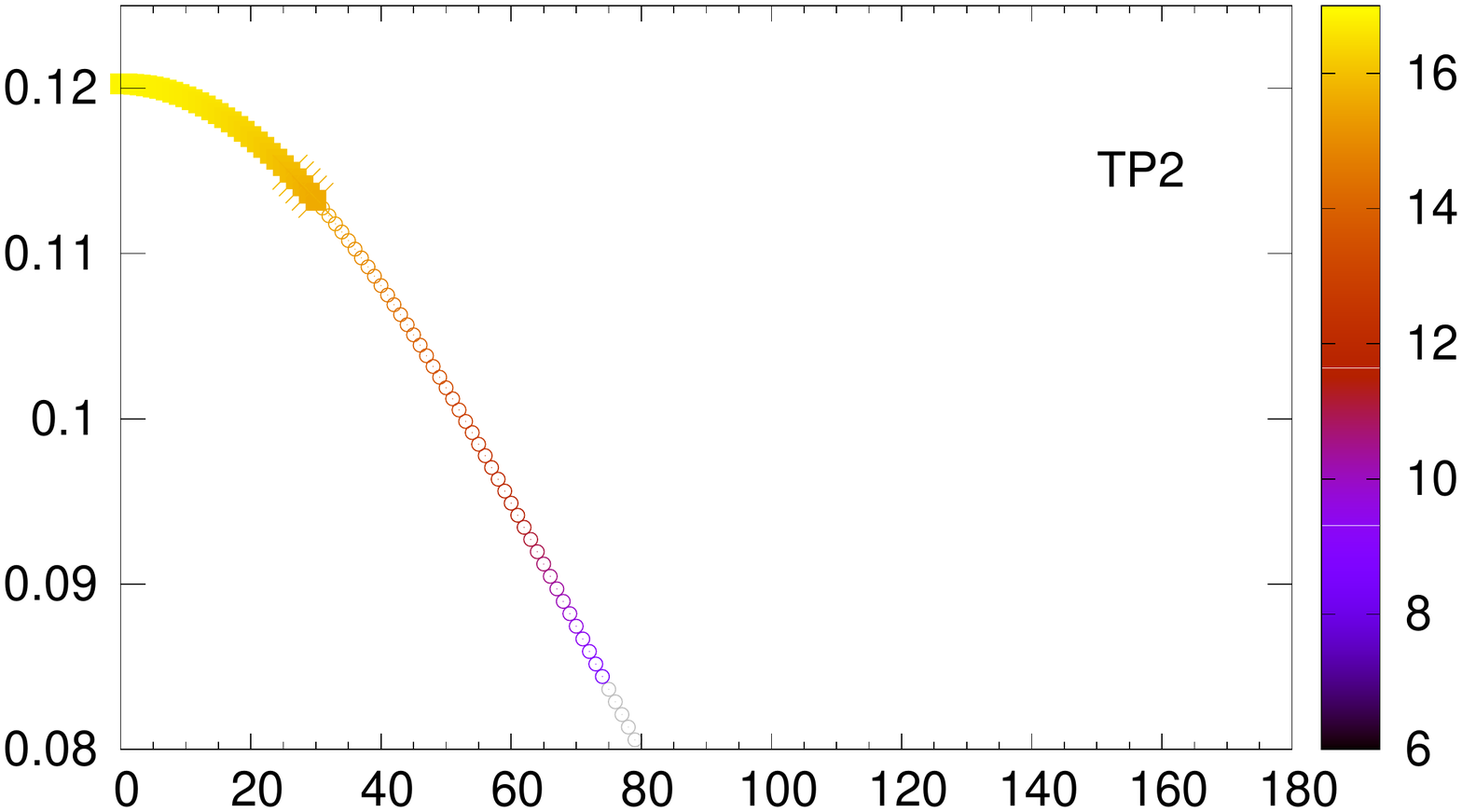} &
\hspace*{-1.3cm}\includegraphics*[width=9.5cm]{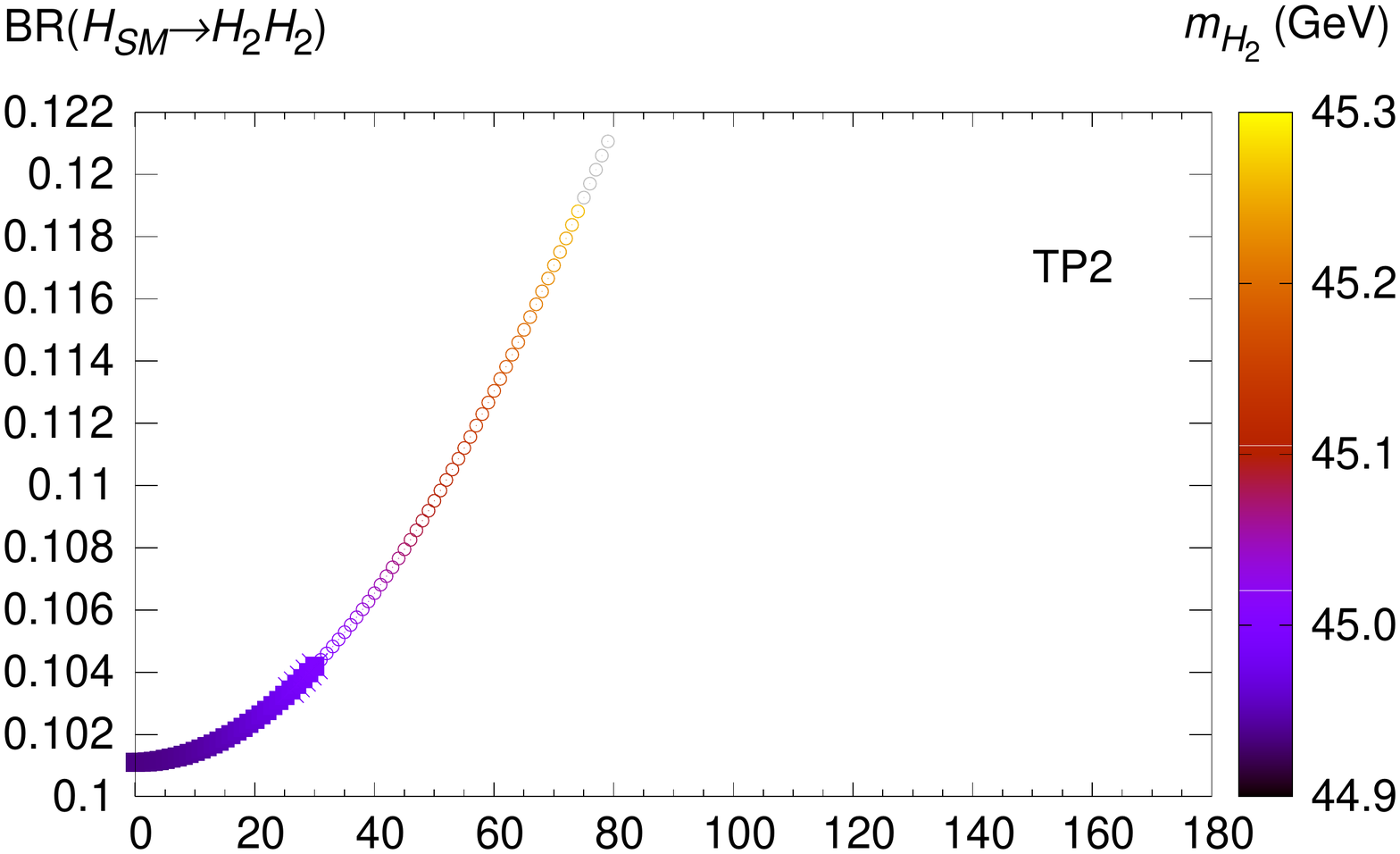} \\
\vspace*{-2.2cm}\hspace*{-1.0cm}\includegraphics*[width=9.5cm]{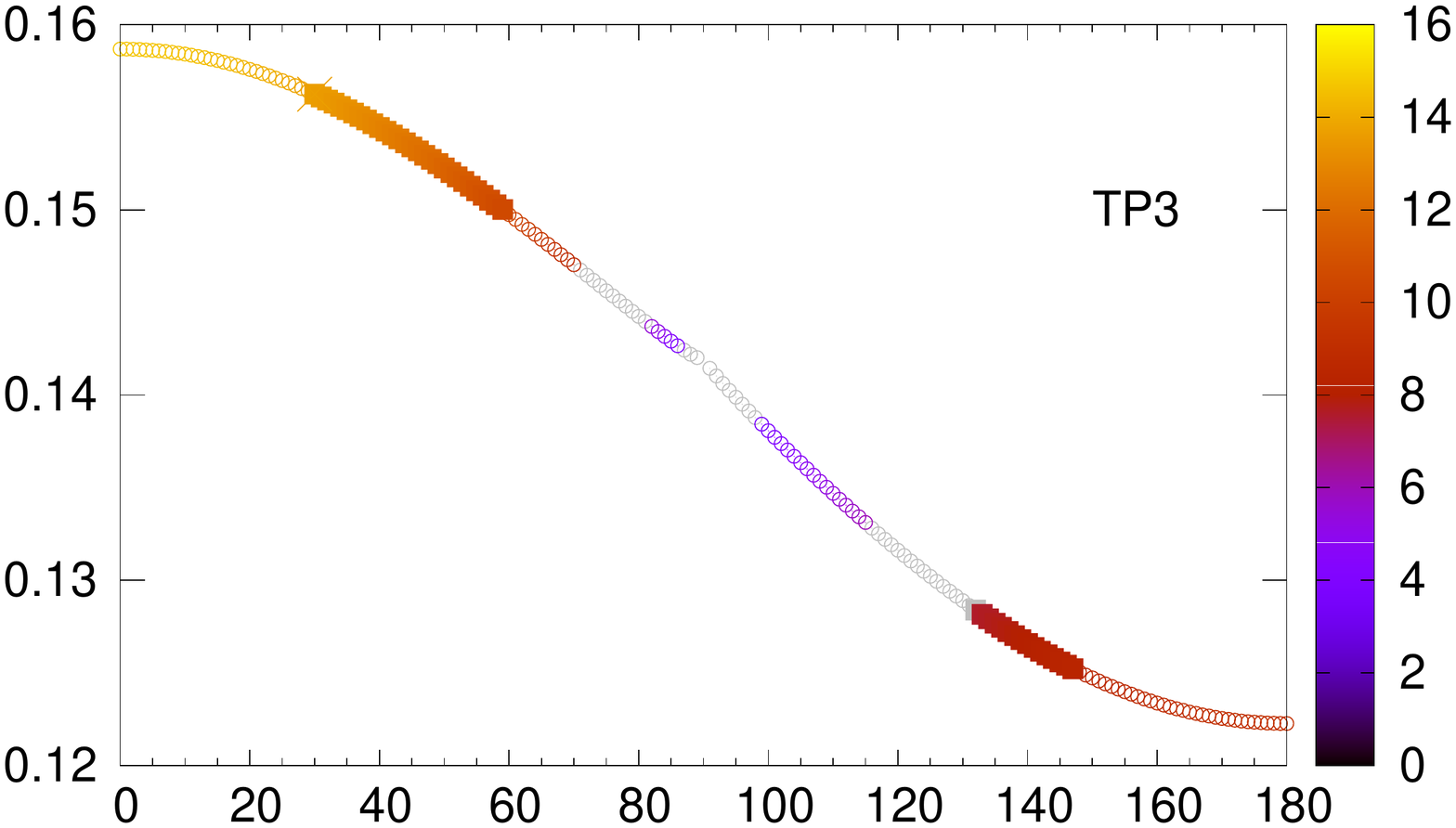} & \\
\vspace*{-1.0cm}\hspace*{-1.0cm}\includegraphics*[width=9.5cm]{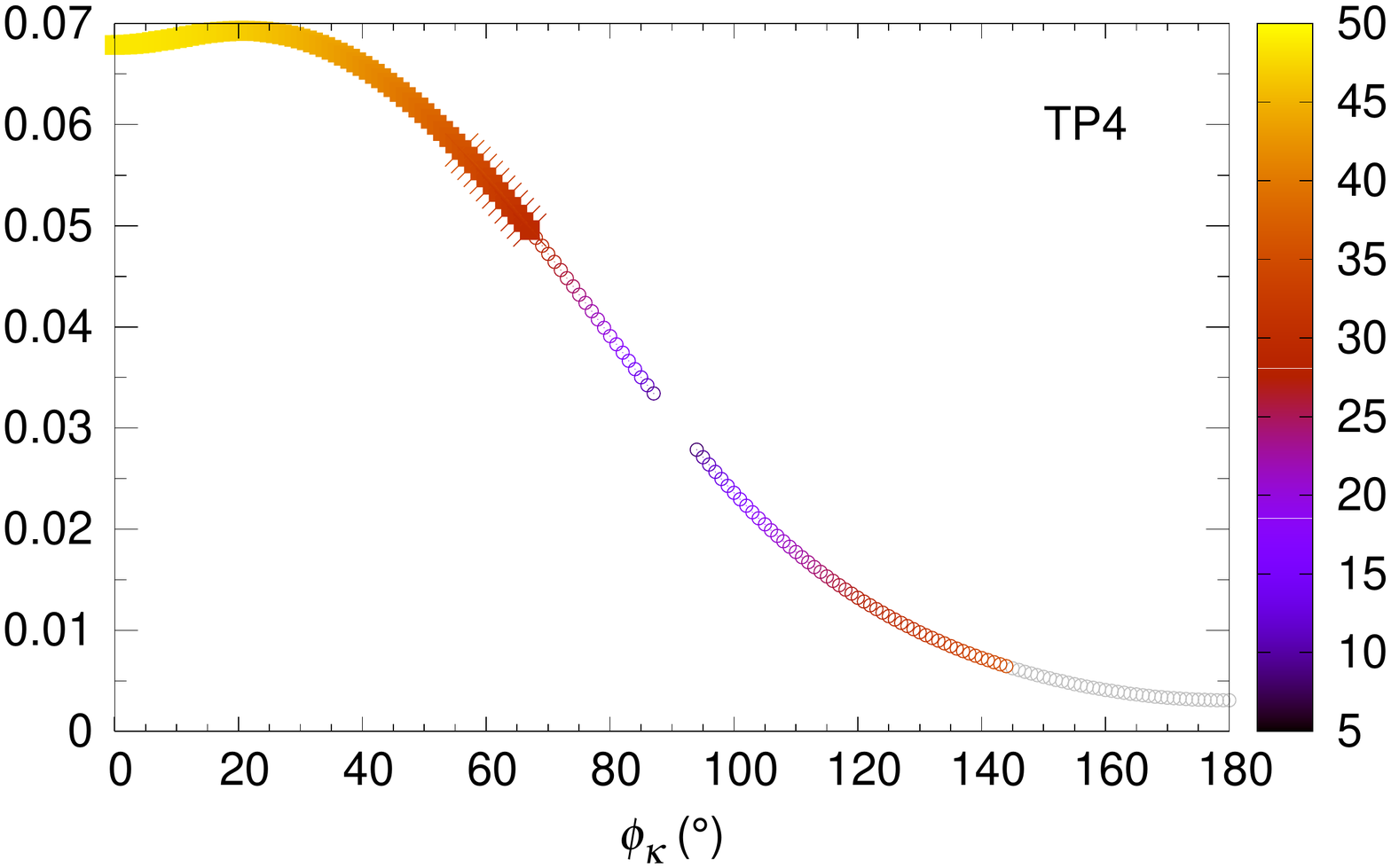} &
\hspace*{-1.3cm}\includegraphics*[width=9.5cm]{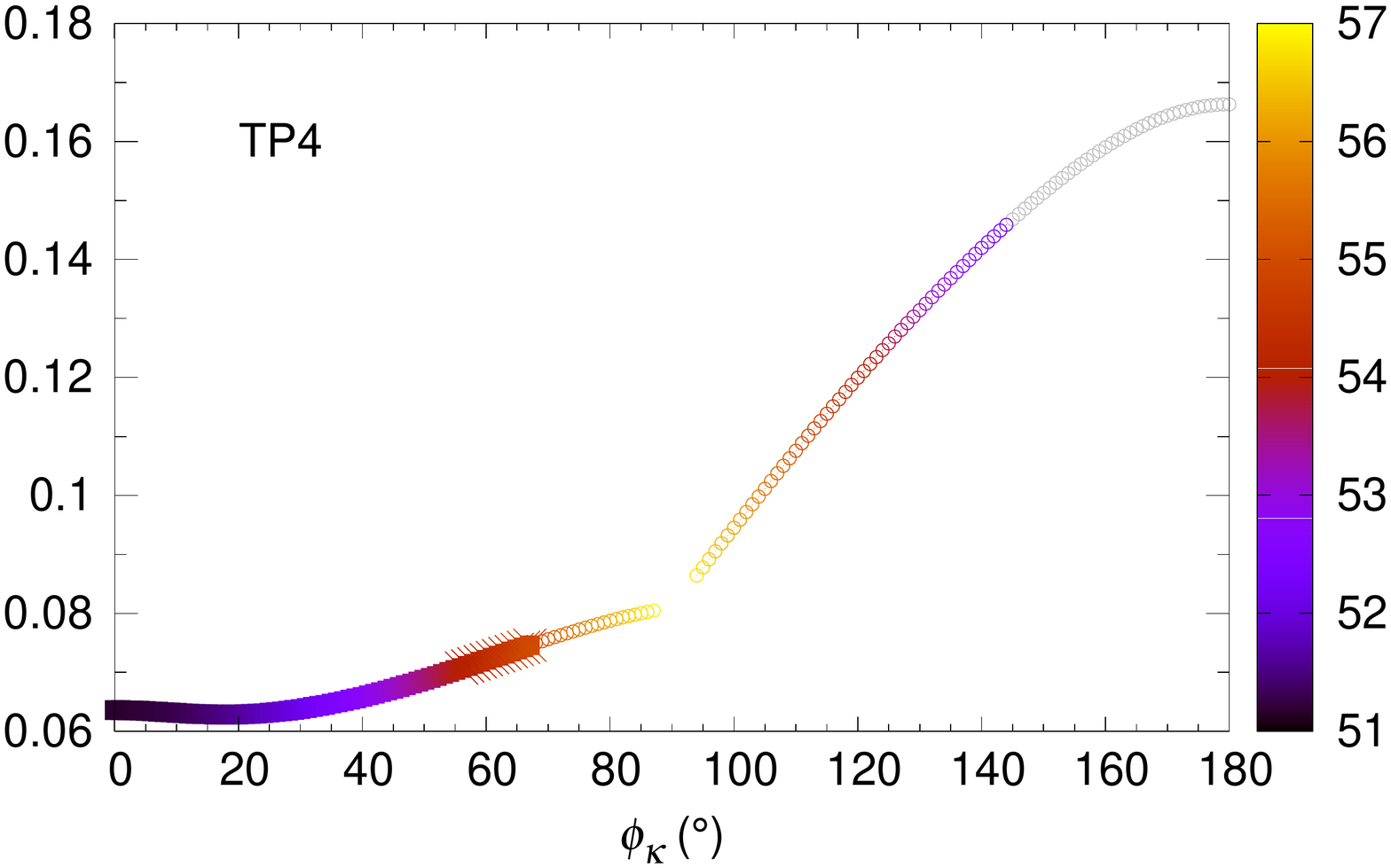} \\
\end{tabular}
\caption{\label{fig:BPs-BRs} BR($\hsm \to H_1H_1$) (left column),  (central column) and BR($\hsm \to H_2H_2$) (right column) as functions of $\phi_\kappa$ for the TPs 1 (top row) -- 4 (bottom row). The heat maps in the left and right columns correspond to $m_{H_1}$ and $m_{H_2}$, respectively. The colouring scheme is the same as in Fig.~\ref{fig:BPs-masses}.}
\end{figure}


\section{\label{sec:concl} Summary and conclusions}
The Higgs sector of the NMSSM can accommodate explicit CP-violating phases at the tree level, whereas in the MSSM such phases enter the Higgs potential only at the higher orders. While the measurements of the leptonic EDMs tightly bound the MSSM-like phase in $A_{\tilde{f}}$, radiatively induced from the sfermion sector, they have been previously found to be much less constraining of the phase of $\kappa$. Importantly, this phase also appears in the tree-level mass term, $2\kappa s$, corresponding to the singlino interaction eigenstate. Therefore, if the $\neut{1}$ is singlino-dominated, its relic abundance can have a strong dependence on $\phi_\kappa$.

The cNMSSM contains 5 neutral CP-indefinite Higgs bosons in total, and any one (or more) of the three lightest of these can fulfil the role of the $\hsm$, in specific regions of the model's parameter space. In this study, we have analysed in detail the quantitative impact of the variation in $\phi_\kappa$ on $\Omega_{\neut{1}} h^2$, for scenarios wherein either $\hsm=H_2$ or $\hsm=H_3$. The $H_1$ was required to always be lighter than 125\,GeV to increase the prospects of self-annihilation of the singlino-like $\neut{1}$ solutions with mass $\lesssim 100$\,GeV, which was our main focus. For certain select values of $\phi_\kappa$, we performed numerical scans of the free parameters of the EW-scale cNMSSM, to find points consistent with a variety of recent experimental constraints, in particular the electron EDM. 

In the overall picture that emerges from this analysis, for specific values of $\phi_\kappa$, nearly exact consistency with the Planck measurement of the DM relic abundance of the Universe is seen for certain $m_{\neut{1}}$ that are precluded in the real NMSSM. Thus, while a large gap appears near $\Omega_{\neut{1}} h^2~0.119$ for $m_{\neut{1}}\sim 10 - 30$\,GeV for $\phi_\kappa = 0^\circ$, this mass range starts filling up as the CP-violation increases, and gets almost entirely covered for $\phi_\kappa \sim 135^\circ$. Evidently, this results from the subtle tweaks in the composition of $\neut{1}$, so that its couplings allow just the right amount of its self-annihilation via one of the multiple potentially resonant sources available in this model, when the CP is violated. 

This inference was confirmed by a closer investigation of the four tests points selected out of the successful points from the scans. For each of these points we varied $\phi_\kappa$ over the entire $0^\circ - 180^\circ$, while fixing the other nine free parameters to their original values. This demonstrated how the different values of $\phi_\kappa$ modify $m_{H_1}$ and $m_{H_2}$, besides the mass as well as the $N_s$ of the $\neut{1}$, to impact the consistency of these points not only with $\Omega h^2$ but also with other experimental data. Furthermore, magnitudes of observables like the BR($\hsm \to \neut{1}\neut{1}$), the BR($\hsm \to H_1 H_1$) and the BR($\hsm \to H_2 H_2$) also show a dependence on $\phi_\kappa$ significant enough that their dedicated inspection might help identify signatures of CP-violation in the NMSSM at the LHC.  


\section*{Acknowledgments}

The authors thank Shabbar Raza for his collaboration in the early stages of this project. MDG acknowledges support from the \mbox{``HiggsAutomator''} (ANR-15-CE31-0002) and \mbox{``DMwithLLPatLHC''} grants of the Agence Nationale de la Recherche (ANR).


\providecommand{\href}[2]{#2}\begingroup\raggedright\endgroup

\end{document}